\documentclass[preprint,nofootinbib]{revtex4}
\usepackage{amssymb,amsmath}
\usepackage{graphicx}
\usepackage{array}
\usepackage{ifthen}

\usepackage{figwidth}

\newlength{\mywidth}
\newcommand{\slashed}[1]{\settowidth{\mywidth}{#1}\hspace*{0.5\mywidth}\makebox[0ex][c]{$#1$}\makebox[0ex][c]{$\slash$}\hspace*{0.5\mywidth}}

\newcommand{\eq}[1]{Eq.~(#1)}
\newcommand{\reffig}[1]{Fig.~\ref{#1}}

\newcommand{\units}[1]{\, \mathrm{#1}}

\newcommand{\ket}[1]{|#1\rangle}

\newcommand{\mean}[1]{\left\langle #1 \right\rangle}

\begin{document}

\title{Axion Cosmology Revisited}

\author{Olivier Wantz}
\email[Electronic address: ]{O.Wantz@damtp.cam.ac.uk}
\affiliation{Department of Applied Mathematics and Theoretical Physics,
Centre for Mathematical Sciences,\\ University of Cambridge,
Wilberforce Road, Cambridge CB3 0WA, United Kingdom}
\author{E.P.S. Shellard}
\email[Electronic address: ]{E.P.S.Shellard@damtp.cam.ac.uk}
\affiliation{Department of Applied Mathematics and Theoretical Physics,
Centre for Mathematical Sciences,\\ University of Cambridge,
Wilberforce Road, Cambridge CB3 0WA, United Kingdom}


\bibliographystyle{plain}

\begin{abstract}
The misalignment mechanism for axion production depends on the temperature-dependent axion mass. The latter has recently been determined within the interacting instanton liquid model (IILM), and provides for the first time a well-motivated axion mass for all temperatures. We reexamine the constraints placed on the axion parameter space in the light of this new mass function. Taking this mass at face value, we find an accurate and updated constraint $ f_a \le 2.8(\pm2)\times 10^{11}\units{GeV}$ or $m_a \ge 21(\pm2) \units{\mu eV}$ from the misalignment mechanism in the classic axion window (thermal scenario). However, this is superseded by axion string radiation which leads to $ f_a \lesssim 3.2^{+4}_{-2} \times 10^{10} \units{GeV}$ or $m_a \gtrsim 0.20 ^{+0.2}_{-0.1} \units{meV}$. In this analysis, we take care to precisely compute the effective degrees of freedom and, to fill a gap in the literature, we present accurate fitting formulas. We solve the evolution equations exactly, and find that analytic results used to date generally underestimate the full numerical solution by a factor $2-3$. In the inflationary scenario, axions induce isocurvature fluctuations and constrain the allowed inflationary scale $H_I$. Taking anharmonic effects into account, we show that these bounds are actually weaker than previously computed. Considering the fine-tuning issue of the misalignment angle in the whole of the anthropic window, we derive new bounds which open up the inflationary window near $\theta_a \to \pi$. In particular, we find that inflationary dark matter axions can have masses as high as 0.01--1$\units{meV}$, covering the whole thermal axion range, with values of $H_I$ up to $10^9$GeV. Quantum fluctuations during inflation exclude dominant dark matter axions with masses above $m_a\lesssim 1$meV.
\end{abstract}

\maketitle

\section{Introduction}
\label{sec:introduction}

Axions are still one of the best motivated cold dark matter candidates. Initially invented to solve the strong CP problem (``why is the QCD vacuum angle so so small?'', i.e.\ $\theta<10^{-9}$), it was soon realised by Weinberg \cite{weinberg:axion} and Wilczek \cite{wilczek:axion} that the Peccei-Quinn (PQ) mechanism \cite{peccei:quinn:cp1,peccei:quinn:cp2} gave rise to a very-light pseudo-scalar Goldstone boson. In order to retain renormalisability, Peccei and Quinn introduced a new chiral symmetry, $U(1)_{PQ}$, on the quark and Higgs fields, that is spontaneously broken. This implies the existence of a new particle, a would-be pseudo-Goldstone boson, the axion; it receives a mass due to instantons because $U(1)_{PQ}$ is anomalous. In the original papers, the axion was incorporated in the electroweak sector but laboratory experiments soon ruled out such a light boson with $\units{GeV}$ coupling. This gave rise to the so-called invisible axion models \cite{shifman:vainshtein:zakharov:cp,kim:axion,dine:fischler:srednicki:axion,zhitnitsky:axion}, that a priori are not tied to any known energy scale. To constrain them, it was realised that such extremely weakly interacting particles could provide a new cooling mechanism for stars. The invisible axions have typically very weak couplings to ordinary matter. On the one hand, this makes their experimental detection difficult but, on the other hand, provides us with a well-motivated dark matter candidate. Refer to past reviews \cite{kim:report,cheng:strongCP:report,turner:axion:report} for further details.

The axion has a rich phenomenology in that it can be produced thermally or non-thermally. The thermal production channel is the standard scenario for most WIMP's \cite{turner:kolb:cosmology}. Recently it was shown that the thermal axion cannot contribute the dominant dark matter component of the universe \cite{hannestad:mirizzi:raffelt:thermal:axion}\footnote{The thermal axion bound follows from $(m_a^{th}/130\units{eV}) < \Omega_c\, g^{dec}_{*}/10$, and is saturated for $m_a^{th} \approx 15 \units{eV}$. This bound is, however, excluded by the new astrophysics bound $m_a<0.01\units{eV}$. Thus, $\Omega_a^{th}=\Omega_c(0.01/15)\approx 0.001 \,\Omega_c$.}. The axion can also be produced non-thermally: after the spontaneous breaking of the PQ symmetry, the axion lives in a $U(1)$ vacuum manifold; such a broken field supports the formation of topological strings \cite{davis:axion:string:1985,davis:axion:string:1986,davis:shellard:axion:string,dabholkar:quashnock:axion:string,battye:shellard:axion:string:1994b} \cite{harari:sikivie:axion:string,hagmann:sikivie:axion:string,hagmann:chang:sikivie:axion:string}, whose radiation produce axions. Finally, axions can be produced non-thermally through the so-called misalignment mechanism: at the QCD phase transition non-perturbative effects generate a mass, and the axion field relaxes to its minimum, which is precisely the PQ mechanism\footnote{The $\theta$ angle, a free parameter, is replaced by a dynamical field that evolves to its CP-conserving minimum.}, invented to solve the strong CP problem. The oscillation around its minimum produce a coherent state of zero mode axions, i.e.\ a Bose-Einstein condensate \cite{sikivie:yang:axion}. This last production scenario is potentially sensitive to the QCD effects, i.e.\ the axion mass, and is the primary subject of this paper.

Because of the anomalous $U_\mathrm{PQ}(1)$ symmetry, the axion has a two gauge boson interaction and can thus decay into two photons; such processes are used to look for axions experimentally, e.g.\ in solar axion searches and vacuum birefringence experiments. The former is one of the more stringent astrophysical constraints, the strongest coming from the analysis of the supernova 1987A neutrino flux which would be affected by axions. It gives a lower bound for the axion decay constant, $f_a \gtrsim 10^9 \mathrm{GeV}$. See \cite{kuster:raffelt:beltran:axions} for a recent, comprehensive set of review articles.

In section \ref{sec:CP:PQ} we briefly review the strong CP problem and the PQ mechanism that lead to the introduction of the axion. We continue to discuss the effective axion potential in section \ref{sec:axion:potential}, and essentially focus on the temperature dependent axion mass. We review our determination of the mass in the framework of the interacting instanton liquid model (IILM) \cite{wantz:iilm:3}. In section \ref{sec:axion:cosmology} we reexamine the cosmology of the vacuum realignment production mode in light of this new mass function: We solve the cosmological evolution equations numerically, and compare the results to the standard analytic approximation, identifying regimes in which present estimates are and are not robust. To make the numerics self-contained, we include the correct effective degrees of freedom for the entropy and radiation density and provide accurate fitting formulas for their rapid evaluation in appendix \ref{app:geff:fit}. For completeness, we also review and update constraints from axion string radiation.

\section{Strong $\mathbf{CP}$ problem and Peccei Quinn Mechanism}
\label{sec:CP:PQ}

The Lagrangian of QCD takes the form
\begin{equation}
\mathcal{L}_{QCD}=-\frac{1}{4} F_{\mu\nu}^{a} F^{a\mu\nu} +i\bar{\psi}(\slashed{D}-M)\psi\, ,
\end{equation}
where $\psi$ carries flavour and colour indices, $M$ is the mass matrix and $D_\mu$ is the covariant derivative in the fundamental representation. If $M=0$ this Lagrangian has a $U(N_F)_L \otimes U(N_F)_R$ symmetry. It turns out that the subgroup $SU(N_F)_L \otimes SU(N_F)_R$ is spontaneously broken down to the vector $SU(N_F)$ symmetry; the Goldstone bosons that accompany the spontaneous symmetry breaking are identified with the light pseudo scalar octet.

Of the remaining phases $U(1)_V \otimes U(1)_A$, the former is conserved through gauge invariance, while the latter is anomalous, i.e.\ is not a symmetry at the quantum level. This follows from the topologically non-trivial nature of QCD and the famous triangle anomaly \cite{pokorski:gauge}.

The topologically non-trivial field space leads to the so-called $n$-vacua, i.e.\ vacua that belong to different topological equivalence classes labelled by a winding number or Chern-Simons charge. The $n$-vacua change their winding under large gauge transformations, $\ket{n} \to \ket{n+m}$ for gauge transformations in the $m$-sector. The winding for the gauge transformation is different from the winding of the $n$-vacua; they are, of course, related in that the winding number for the non-vacuum field configurations is given by the divergence of the Chern-Simons current. Thus to construct a gauge invariant vacuum, we need a superposition of the $n$-vacua, and we arrive at the $\theta$-vacua
\begin{equation}
\ket{\theta}=\sum_n e^{i\theta n} \ket{n}\, .
\end{equation}
Due to a superselection rule, different values of $\theta$ correspond to different theories. Summing the partition function over all topological sectors introduces a new term into the action, and the Lagrangian becomes
\begin{equation}
\mathcal{L} \to \mathcal{L}+\theta\frac{g^2}{32\pi^2}F_{\mu\nu}^{a} \tilde{F}^{a\mu\nu}\, . \label{Leff}
\end{equation}
The $\theta$ term violates P and CP unless $\theta=0$ or $\theta=\pi$. It gives an electric dipole moment to the neutron \cite{baluni:cp:violation:qcd,crewther:divecchia:veneziano:witten:electric:dipole:moment}, which is tightly constrained experimentally \cite{baker:electric:dipole:neutron} and results in the bound
\begin{equation}
\theta < 10^{-9}\, .
\end{equation}

A priori, we could just demand that the QCD Lagrangian be CP symmetric. However, in the Standard Model of particle physics the electro-weak sector contributes to the vacuum angle through the phase of the quark mass matrix; the weak CP non-invariance is well accommodated for by the CKM-matrix, and thus by explicit CP breaking. Therefore, it would be rather unnatural to postulate a CP invariant QCD Lagrangian, and CP is presumably also broken explicitly in the strong sector.

The strong CP problem relates to explaining the smallness of the $\theta$ angle. A solution within QCD is given if at least one quark is massless. In such a case, the massless quark can be chirally rotated by $\theta$ which, through the anomaly, removes the $\theta$ term altogether from the Lagrangian. Hence, the $\theta$ parameter is not physical. However, the assumption of a massless quark doesn't comply to experimental observations combined with predictions from chiral perturbation theory\footnote{The non-zero quark mass used in chiral perturbation theory could in fact be an instanton induced `soft' mass, even though the fundamental current mass is zero, see for instance the discussion in \cite{kim:axion:quintessential,kim:axion:cdm}. Recent lattice studies to determine the current quark masses seem to rule this possibility out \cite{mason:trottier:horgan:davies:lepage:quark:masses}.}, see for instance \cite{weinberg:qft}.

All models have the generic feature that the axion only couples derivatively to matter and the only non-derivative coupling is to the topological charge\footnote{If we restrict ourselves to QCD. The axial current does, however, also receive an anomaly from QED; the corresponding non-derivative interaction is responsible for the decay of an axion into two photons. It is exactly the same mechanism that explains why the neutral pion decays into two photons.}
\begin{equation}
\mathcal{L}^{non-der.}_a = i\frac{\phi_a}{f_a} \frac{g^2}{32\pi^2} F^{a}_{\mu\nu} \tilde{F}^{a}_{\mu\nu} \,,
\end{equation}
where $\phi_a$ is the axion field and $f_a$ the axion decay constant.

The PQ mechanism works because the effective potential for the (homogeneous) axion field has a CP-conserving minimum \cite{kim:report}
\begin{eqnarray}
e^{-V  V_\mathrm{eff}(\phi)} &=&\left|\int [dA] \det(\gamma_\mu D_\mu + M)\,e^{-S+i(\theta+\frac{\phi}{f_a})\mathcal{Q}} \right|\,, \label{eq:axion:effective:potential}\\
&\le& \int [dA] \det(\gamma_\mu D_\mu + M)\, e^{-S}\left|e^{i(\theta+\frac{\phi}{f_a})\mathcal{Q}} \right|\,, \\
&=& e^{-V V_\mathrm{eff}(\phi=-f_a \theta)}\, ,
\end{eqnarray}
and thus $\mean{\theta+\phi_a/f_a}=0$, where $V$ the 4-dimensional volume. Note that the effective potential is periodic and that our computation is performed in the Euclidean theory. At finite temperature, real-time configurations, sphaleron transitions, might also give contributions. However, as was shown in \cite{mclerran:mottola:axion:sphaleron}, these classical field configurations do not affect the axion mass at leading order, basically because the classical field equations do not depend on $\theta$. Shifting the axion field, we will define $\theta + \phi/f_a \to \theta_a$, with $\theta_a$ the axion angle. Thus, the PQ mechanism effectively trades $\theta$, a free parameter, for a dynamical field that evolves to its CP-conserving minimum.

\section{Axion potential}
\label{sec:axion:potential}

Evaluation of (\ref{eq:axion:effective:potential}) allows us, in principle, to determine the axion effective potential. We can of course not hope to get exact, analytic formulas since the effective potential includes strong coupling QCD effects. Ultimately, the lattice will be able to compute the axion effective potential exactly. However, there are technical problems since the action is complex and cannot be studied directly by simple Monte Carlo methods. This is the same problem that arises in lattice gauge theory with a finite baryon density, i.e.\ a non-zero chemical potential.

On the analytic front, progress can be made by saturating the path integral with suitable background configurations that are supposed to play an important role for the problem at hand. Since the axion couples to the topological charge, it makes sense to study the path integral in the background of instantons, which are the prototype topological field configurations in QCD. This approach was pioneered in \cite{callan:dashen:gross:theory:strong:interactions,callan:dashen:gross:theory:hadronic} with a strong emphasis on the so-called dilute gas approximation. In this limit, the partition function is saturated by an ensemble of non-interacting instantons, and the path integral simply factorises into single instanton contributions. The latter has been computed exactly, at zero \cite{thooft:instanton:fluctuations} and finite temperature \cite{gross:pisarski:yaffe:instantons:finite:T}, and is of the form (at one-loop and zero temperature)
\begin{eqnarray}
Z_1 &=& \int d\rho d(\rho) \prod_{q=1}^{N_F}\det(\slashed{D} + m_q)\,,\\
d(\rho) &=& \rho^{b-5} \Lambda^b \left(\frac{8\pi^2}{g^2}\right)^{2N_c} C_{N_c}\,,\\
b &=& \frac{11}{3} N_c -\frac{2}{3}N_F\,,\\
C_{N_c} &=& \frac{\displaystyle 0.466\exp(-1.679N_c)}{\displaystyle (N_c-1)!(N_c-2)!}\,.
\end{eqnarray}
Note that $\det$ stands for the renormalised quark determinant; its UV contribution gives rise, through charge renormalisation, to the $N_F$ dependent term in $d$. Recently, the quark determinant has been computed for any value of the quark masses \cite{dunne:hur:lee:min:instanton:determinant:mass}; in the small mass limit we have $\det(\slashed{D} + m_q) \to 1.34 m_q \rho$. At finite temperature, electric Debye screening prohibits large scale coherent field configurations to exist in the plasma, i.e.\ fields with a correlation length $\xi \gtrsim 1/T$. This changes the instanton weight by $d(\rho) \to d(\rho) \exp(-(2\pi T \rho)^2)$.

Given $Z_1$, the dilute gas approximation follows immediately as
\begin{equation}
 Z = \sum_{N_I,N_A}\frac{1}{N_I!N_A!} Z_1^{N_I+N_A} \exp(i\theta_a(N_I - N_A))\,,
\end{equation}
where we have already included the axion angle through its non-derivative coupling to the topological charge, which is approximated by $Q=N_I-N_A$ in the background of $N_I$ instantons and $N_A$ anti-instantons. This sum is easily computed, and we find the effective potential
\begin{equation}
 V_\mathrm{eff}(\theta_a)=-2\int d(\rho) \cos\theta_a\,,
\end{equation}
where we absorbed the renormalised determinant into $d$ for notational simplicity. That the potential had to be periodic follows from the defining QCD path integral (\ref{eq:axion:effective:potential}). In general, the potential will have a much more complicated form than this simple cosine, see \cite{halperin:zhitnitsky:yang:mills:theta,halperin:zhitnitsky:qcd:theta,halperin:zhitnitsky:qcd:axion:potential,fugleberg:halperin:zhitnitsky:domain:walls:theta:qcd,gabadadze:shifman:vacuum:structure:qcd,gabadadze:shifman:qcd:vacuum:axion}. Still, one can estimate that the first few terms of a harmonic expansion should describe the axion potential rather accurately \cite{kim:report}, and it is custom for axion cosmology to work with a simple cosine.

Note that the effective potential is negative at its CP conserving minimum. The reason is that the one-instanton contribution is really normalised to the perturbative result, which by definition has zero energy. Instantons, interpreted as tunnelling effects between the $n$-vacua, restructure the vacuum and give rise to the energetically lower true $\theta$-vacuum.

At temperatures below the QCD phase transition, the dilute gas approximation breaks down. In that regime chiral perturbation theory can however be used to derive the classic result for the zero temperature axion mass
\begin{equation}
 m^2_a f^2_a = m^2_\pi f^2_\pi \frac{m_u m_d}{(m_u+m_d)^2}\,, \label{eq:axion:mass:chiral_theory:T0}
\end{equation}
with corrections of order one that depend on the precise model. The above axion mass is exact in the framework of the simplest hadronic axion \cite{shifman:vainshtein:zakharov:cp,kim:axion}, where only a new super-heavy $SU(2) \times U(1)$ singlet Dirac spinor carries PQ charge; after integrating out the heavy field, we are left with the typical non-derivative coupling to the QCD topological charge. The low energy effective QCD Lagrangian, i.e.\ chiral perturbation theory, including the axion, is also used to derive the couplings of axions with pions and the weak sector, needed in the thermal production scenario \cite{hannestad:mirizzi:raffelt:thermal:axion}. Above the phase transition these coupling follow from the fundamental Lagrangian \cite{masso:rota:zsembinszki:axion:thermal}.

This same result can also be derived from the defining QCD path integral by noting that
\begin{equation}
 m^2_a f^2_a = \frac{\partial^2 V_\mathrm{eff}}{\partial \theta_a^2} \equiv \chi\,, \label{eq:axion:mass:T0}
\end{equation}
where $\chi = \lim_{V\to\infty} \frac{\langle Q^2 \rangle}{V}$ is the topological susceptibility. It can be shown that $\chi$ is related to the quark condensate \cite{leutwyler:smilga:spectrum:dirac} and, using chiral perturbation theory, this can be transformed into (\ref{eq:axion:mass:chiral_theory:T0}). Here, only the axion carries PQ charge and it interacts with matter derivatively, together with the topological coupling to the gauge sector. This axion field is the same as the physical axion field in the effective Lagrangian approach, i.e.\ the propagation eigenstate.

The important point to note is that the axion mass is essentially given by the topological susceptibility, a quantity that is routinely measured on the lattice. With the recent progress of lattice algorithms and increased computing power, physical quark mass simulations are finally feasible and the lattice will soon be able to provide us with the best estimate for the temperature dependent axion mass.

In this paper we will use the interacting instanton liquid model (IILM) to study the axion mass; it is based on the idea that the instanton ensemble is fairly dilute but that interactions are nevertheless important \cite{schafer:shuryak:instantons:qcd:review}. It has been shown to give a self-consistent framework at zero and finite temperature \cite{diakonov:instanton:variational,diakonov:instanton:quarks,diakonov:instanton:nonzero:T}. The IILM saturates the path integral with an ansatz for the multi-instanton background configurations, and treats the low-frequency fluctuations `exactly' while still assuming a factorisation of the high frequency gluon, quark and ghost spectrum. The IILM partition function is defined by
\begin{eqnarray}
Z_\mathrm{IILM} &=& \sum_{N_I,N_A} \frac{1}{N_I!N_A!} \int \prod_{i=1}^{N_I+N_A} d\gamma_i d(\rho_i) e^{-S_\mathrm{int}}\,,\\
S_\mathrm{int} &=& \sum_{\mathrm{pairs}\,(i,j)} S_0(\sqrt{\rho_i \rho_j}) V_{ij} - \sum_{q=1}^{N_F}
 \left\{
 \begin{array}{cl}
 \ln\det(\mathbb{I} + \frac{TT^\dagger}{m^2_q}) & , Q<0 \\
 \ln\det(\mathbb{I} + \frac{T^\dagger T}{m^2_q}) & , Q>0
 \end{array}
\right. .
\end{eqnarray}
The integration is over the collective coordinates, which are the positions, sizes and the colour embedding matrices. The classical gluonic two-body interaction is given by $V_{ij}$; it receives a contribution from the high frequency fluctuations through charge renormalisation. The latter is approximated by the one-instanton action, $S_0=8\pi/g^2$, with the running coupling evaluated at $\sqrt{\rho_i \rho_j}$. The low frequency quark determinant is approximated by the finite dimensional subspace of quasi zero modes $\{\xi_n\}$, i.e.\ $T_{IA}=\langle \xi_I|\slashed{D}|\xi_A \rangle$.

\begin{figure}[tbp]
\begin{center}
 \includegraphics[width=\figwidth,clip=true,trim=0mm 0mm 15mm 10mm]{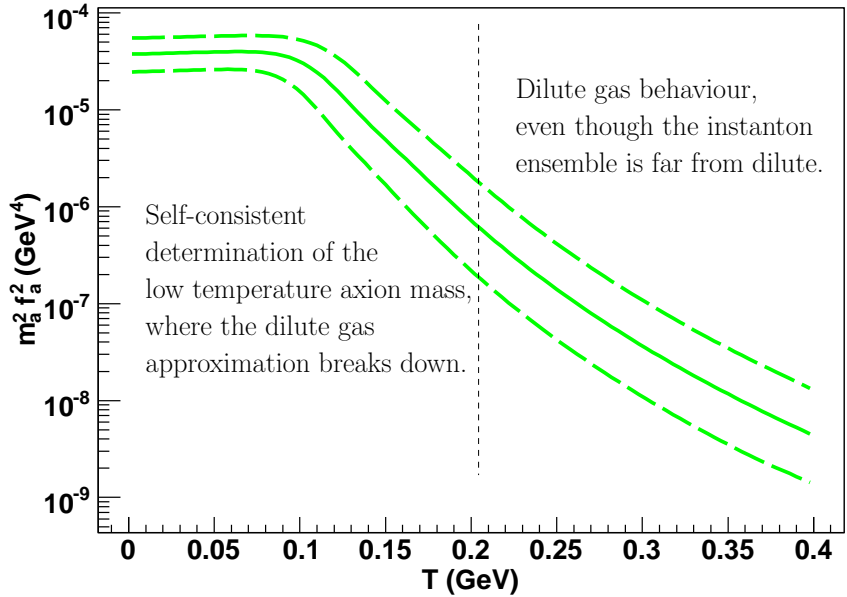}
\end{center}
\caption{The mass for the QCD axion follows from the topological susceptibility, $m^2_a f^2_a = \chi$. The fit goes over to the dilute gas approximation for moderately high temperatures $T\approx 400\units{MeV}$, in accordance with the IILM data. Note that the large errors are mostly due to the large uncertainties in the determination of $\Lambda$, used to set dimensions.}\label{fig:mass}
\end{figure}

This partition function has been used to run grand canonical Monte Carlo simulations. The parameters were calibrated at zero temperature and physical quark masses were determined self-consistently \cite{wantz:iilm:1}. Finite temperature simulations have been performed to compute the topological susceptibility \cite{wantz:iilm:3}. In the latter paper that data was used to derive the axion mass. In the low temperature regime the axion mass can be approximated by
\begin{equation}
m^2_a f^2_a = 1.46\;10^{-3}\Lambda^4 \frac{1+0.50\,T/\Lambda}{\displaystyle 1+\left(3.53\, T/\Lambda\right)^{7.48}},\,  0 < T/\Lambda < 1.125 \,,\label{eq:mass:iilm}
\end{equation}
where $\Lambda=400\units{MeV}$. It is displayed in \reffig{fig:mass}.

Including effects from quark thresholds, the high temperature axion is given by
\begin{equation}
m^2_a f^2_a = \Lambda^4
\left\{
 \begin{array}{l@{,\;}l}
 \exp\left[ d^{(3)}_0 + d^{(3)}_1 \ln\frac{T}{\Lambda} + d^{(3)}_2 \left(\ln\frac{T}{\Lambda}\right)^2 + d^{(3)}_3 \left(\ln\frac{T}{\Lambda}\right)^3 \right] & T^{(3)} < T < T^{(4)}\\
 \exp\left[ d^{(4)}_0 + d^{(4)}_1 \ln\frac{T}{\Lambda} + d^{(4)}_2 \left(\ln\frac{T}{\Lambda}\right)^2 \right] & T^{(4)} < T < T^{(5)}\\
\exp\left[ d^{(5)}_0 + d^{(5)}_1 \ln\frac{T}{\Lambda} + d^{(5)}_2 \left(\ln\frac{T}{\Lambda}\right)^2 \right] & T^{(5)} < T < T^{(6)}\\
\end{array}
\right. ,
\end{equation}
and the different parameters by
\begin{equation}
\begin{array}{c|c|c|c|c|c}
N_f & d^{(N_f)}_0 & d^{(N_f)}_1 & d^{(N_f)}_2 & d^{(N_f)}_3 & T^{(N_f)} \units{GeV} \\\hline\hline
3 & -15.6 & -6.68 & -0.947  & +0.555 & 0.45 \\\hline
4 & +15.4 & -7.04 & -0.139  &   -    & 1.2 \\\hline
5 & -14.8 & -7.47 & -0.0757 &   -    & 4.2 \\\hline
6 &   -   &   -   &   -     &   -    & 100 \\
\end{array} \label{eq:mass:dga:full}
\end{equation}
The quark thresholds are treated within the effective field theory language, where decoupling is enforced by hand and continuity is achieved through matching conditions.

\begin{figure}[tbp]
\begin{center}
 \includegraphics[width=0.95\figwidth,clip=true,trim=0mm 0mm 15mm 10mm]{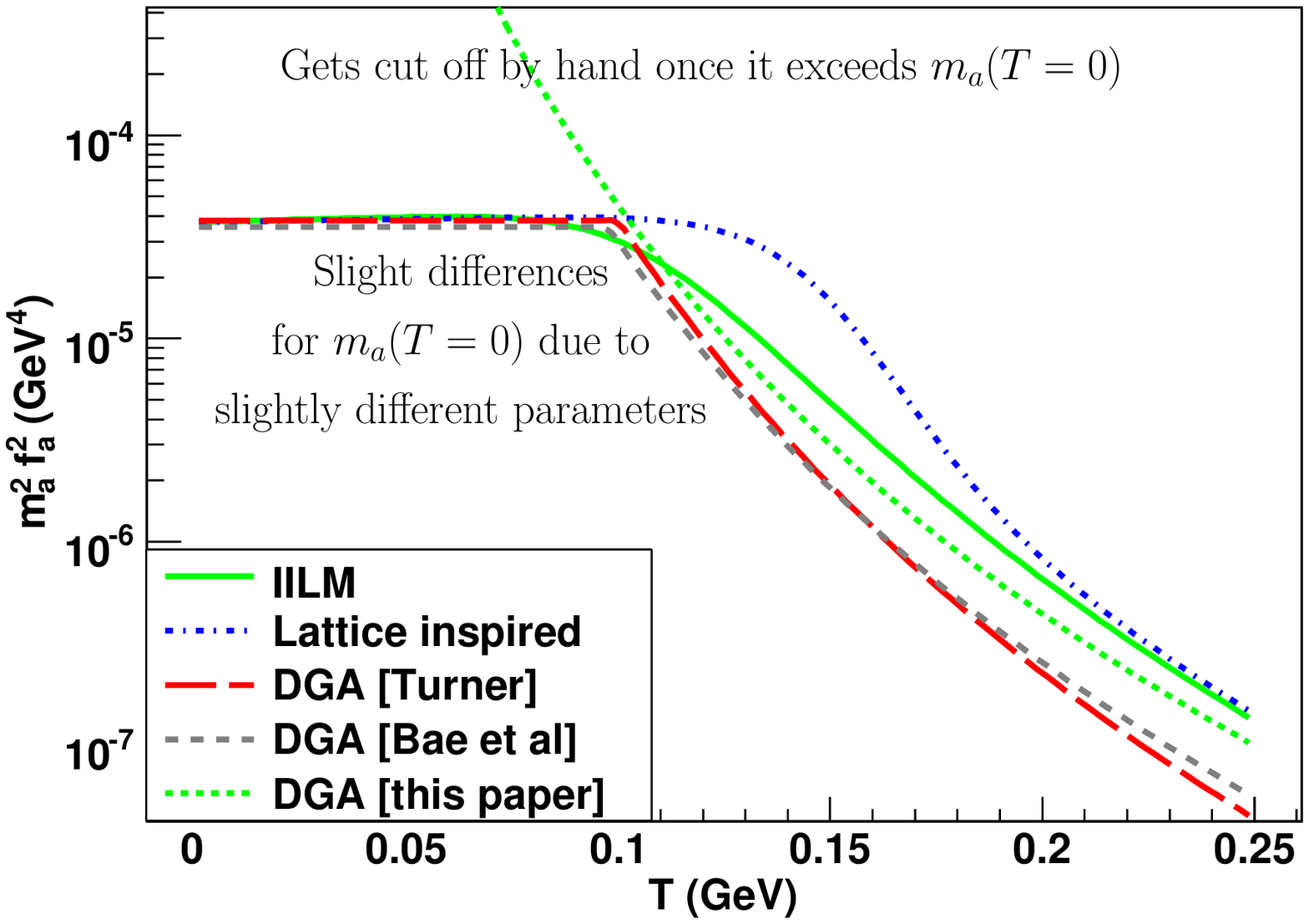}
 \includegraphics[width=0.95\figwidth,clip=true,trim=0mm 0mm 15mm 10mm]{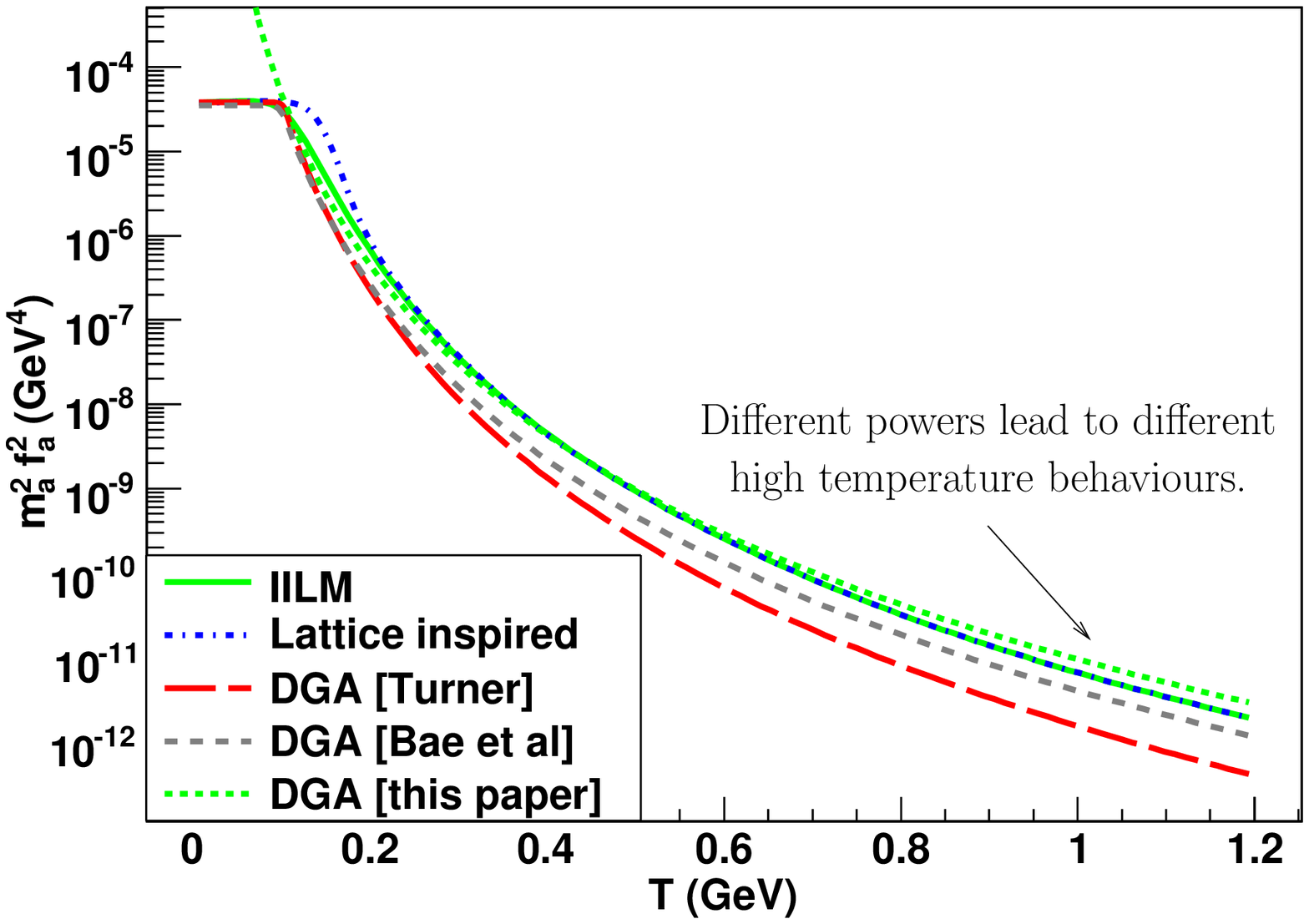}
\end{center}
\caption{Shown are the mass for the QCD axion from IILM simulations (\ref{eq:mass:iilm}), from a lattice inspired fit that uses the IILM mass shifted towards higher temperatures to mimic the phase transition at $T^{lat}_c\approx 160 \units{MeV}$, from the classic dilute gas approximation (DGA) by Turner \cite{turner:axion:cosmology} and its update by Bae et al. \cite{bae:huh:kim:axion}, and from the DGA derived in this paper (\ref{eq:mass:dga}). The simple power-law DGA axion masses are cut off by hand once they exceed $m_a(T=0)$ and give a surprisingly good approximation to the full IILM result; we believe this is a coincidence. The differences that persist to high temperatures, between the update and our DGA model, arise from the slightly different quark masses. Our choice has the merit that the masses were determined self-consistently within the IILM at $T=0$ \cite{wantz:iilm:1}.}\label{fig:masses}
\end{figure}

We also give a very simple approximation to the dilute gas result in the form of a power-law, as in earlier work \cite{turner:axion:cosmology,bae:huh:kim:axion},
\begin{equation}
 m^2_a = \frac{\alpha_a \Lambda^4}{f_a^2 (T/\Lambda)^n}\,,\label{eq:mass:dga}
\end{equation}
where $n=6.68$ and $\alpha=1.68 \,10^{-7}$, from (\ref{eq:mass:dga:full}); it compares well with \cite{bae:huh:kim:axion}. We believe it is a coincidence that such a simple fit, based solely on the high temperature regime, still gives such a good overall approximation to the much more elaborate result of the IILM simulations, see \reffig{fig:masses}.

We found that the instanton ensemble is very distinct from a non-interacting system. Corroborating earlier ideas on the instanton liquid at finite temperature \cite{wantz:iilm:3}, we found a population of instanton--anti-instanton molecules and a non-interacting remnant. The molecules do not lead to charge fluctuations and, hence, the axion mass is determined by the random sub-ensemble. It turns out that the latter have a concentration that just matches the dilute gas approximation. We believe this is an unfortunate coincidence; in particular, we have found within a toy-model that, depending on the interaction and screening effects, a different high temperature behaviour can occur: for stronger interactions the molecule concentration can become higher so that the non-interacting sub-ensemble acquires a lower density, and hence a lower axion mass, compared to the dilute gas estimate \cite{wantz:iilm:3}. A crude argument within the IILM gave evidence that at higher temperatures, with more active quark flavours, the fermionic interactions might outweigh the screening effects and the molecule concentration could increase. For temperatures below the charm or even the bottom threshold, the molecule concentration will, however, decrease as the screening effects dominate over the interactions. That the molecule concentration would not depend monotonically on the temperature seems unnatural. Below the charm threshold, where we find that the molecule density decreases, weaker screening effects could alter this trend: corrections to the factorised high frequency quantum interactions could indeed induce weaker screening because overlapping instantons have effectively a smaller size; note that at zero temperature such quantum interactions were estimated to be subdominant but this has not been repeated for the finite temperature case.

In the pure gauge sector the IILM is not able to accurately describe the topological susceptibility as obtained from lattice simulations, whereas a dilute gas model of non-trivial holonomy calorons fared rather better \cite{gerhold:ilgenfritz:mueller_preussker:kvbll:gas:confinement}. In particular, the IILM predicts a topological susceptibility that decays too fast. These more general degrees of freedom should therefore be included into the IILM in the future to improve on the quenched sector and to investigate whether they also lead to significant changes in the unquenched case. We might expect the implications to be smaller because of chiral symmetry, which is successfully implemented in the IILM and believed to play a crucial role in the strong dynamics.  In the unquenched case, we don't have lattice data available to check for a qualitatively different behaviour of the IILM in the high temperature region. Given the progress of lattice simulations in the physical regime, this issue will be settled in the near future.

Despite these uncertainties, our investigation might be evidence that the axion mass is fairly insensitive to the details of the instanton ensemble and that the dilute gas approximation might prove to be a reliable estimate, even though the instanton ensemble is certainly not non-interacting.

\section{Axion cosmology}
\label{sec:axion:cosmology}

Now that we have determined the mass function for the axion, we can turn to the cosmological implications. For the misalignment mechanism to produce a dominant axion contribution, we assume that the axion is created before the end of inflation; otherwise axionic string radiation will produce the bulk of the axion population \cite{davis:axion:string:1985,davis:axion:string:1986,davis:shellard:axion:string,dabholkar:quashnock:axion:string,battye:shellard:axion:string:1994b}\footnote{Note that there is still some controversy with regard to this statement, and other numerical work \cite{harari:sikivie:axion:string,hagmann:sikivie:axion:string} finds the axion contribution from string radiation and decay to be comparable to the contribution from the vacuum realignment production mode.}. Thus when the PQ symmetry is spontaneously broken, we have different initial angles, the misalignment angles, in the causally disconnected regions. Later, inflation sets in and stretches these patches to cosmological sizes such that throughout the observable universe the same misalignment angle prevails. Once instanton effects set in, the axion becomes massive and starts to oscillate.

Only the homogeneous part of the axion field is relevant for the misalignment mechanism. We are interested in the regime when the axion starts rolling, which happens around the QCD scale when the universe is radiation dominated. Recent lattice data suggests that the QCD phase transition is a cross-over rather than a sharp phase transition \cite{aoki:fodor:katz:szabo:transition:temperature}. In particular, it is not of first order, and no latent heat is produced. Furthermore, we assume that no exotic particles beyond those in the Standard Model decouple relativistically. Thus, the universe evolves adiabatically and we leave aside models with additional entropy production, see for instance the discussion in \cite{fox:pierce:scott:axion:cosmology}.

Since we will also be interested in assessing the accuracy of the standard analytic results for the misalignment mechanism, we will take care to include the correct number of degrees of freedom $g_*$. Following the analysis in \cite{coleman:roos:geff}, we include all the known hadrons\footnote{``with strong experimental evidence'', see \cite{amsler:2008:pdg}; we use a multiplicity of $g_J=1$ if the angular momentum is not known} up to a mass of $3 \units{GeV}$ in the low temperature regime, and match $g_*$ smoothly to the plasma phase. Using a rough estimate of the available lattice data sensitive to the confinement/deconfinement transition \cite{karsch:recent:partII}, e.g.\ the Polyakov loop, we will use $T_c=180\units{MeV}$; given the spread in these results and also the cross-over nature of the transition \cite{aoki:fodor:katz:szabo:transition:temperature}, we apply the smoothing over a range of $\Delta T=20\units{MeV}$. This is slightly different from \cite{coleman:roos:geff} which used $\{T_c=180\units{MeV},\Delta T=5 \units{MeV}\}$ but  the changes are small in any case. The effective degrees of freedom are given by \cite{turner:kolb:cosmology}
\begin{eqnarray}
 g_{*,R} &=&\sum_i \left(\frac{T_i}{T}\right)^4 \frac{15 g_i}{\pi^4} \int_0^\infty dx \frac{\sqrt{x^2+y_i^2}}{\exp\sqrt{x^2+y_i^2}+(-1)^{Q^f_i}}\,, \label{eq:geff:R}\\
 g_{*,S} &=&\sum_i \left(\frac{T_i}{T}\right)^3 \frac{45 g_i}{4 \pi^4} \int_0^\infty dx \frac{x^2 \sqrt{x^2+y_i^2}}{\exp\sqrt{x^2+y_i^2}+(-1)^{Q^f_i}}\left( 1+\frac{1}{3}\frac{x^2}{x^2+y^2_i} \right)\,, \label{eq:geff:S}
\end{eqnarray}
where $T$ is the temperature of the plasma, $T_i$ the temperature of species $i$, $y_i=m_i/T_i$, and $Q^f(\mathrm{fermion})=1$ and $Q^f(\mathrm{boson})=0$. The full numerical integration is too slow to be used in other numerical investgations, such as the axion dynamics in a Friedmann-Robertson-Walker (FRW) universe to be discussed below. To this end, we have also determined fits that are accurate below the $1\%$ level, except at the phase transition and $e^\pm$ annihilation where the error rises briefly to $4\%$. The fits are given in appendix \ref{app:geff:fit}.

\begin{figure}[tbp]
\begin{center}
 \includegraphics[width=\figwidth,clip=true,trim=0mm 0mm 15mm 10mm]{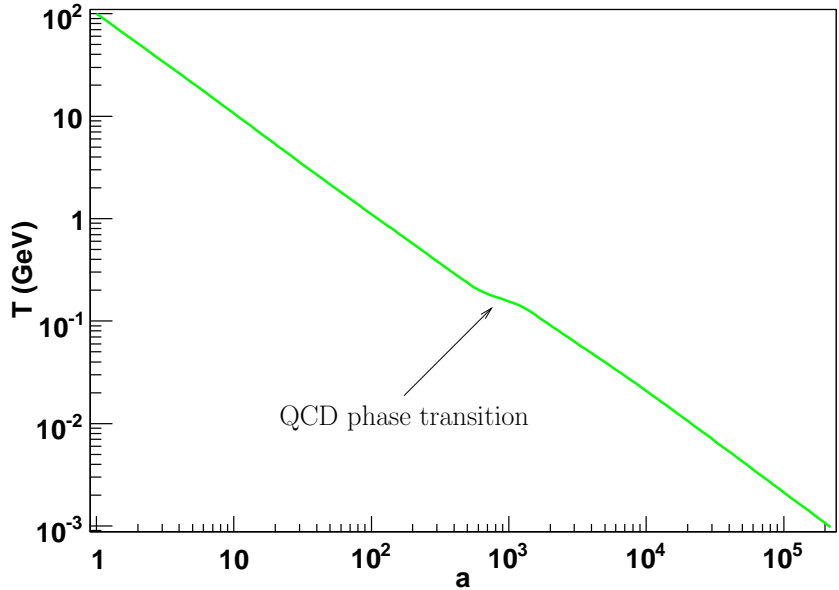}
\end{center}
\caption{In an adiabatically evolving universe the scale factor and the temperature are related through the condition of constant entropy. Given the knowledge of the effective degrees of freedom $g_{*,S}$, it amounts to solving an implicit equation. The QCD phase transition occurs at around $T_\mathrm{QCD}\approx 180 \units{MeV}$, when the number of hadronic excitations rises very sharply, and $g_{*,S}$ is almost discontinuous; the would-be latent heat `reheats' the universe, which is clearly seen in the graph.}\label{fig:aT}
\end{figure}

An adiabatically evolving universe has a specific relation between the temperature and the scale factor, see Fig.~\ref{fig:aT}. This allows us to accurately relate cosmic time to the temperature of the plasma; the latter is required to evaluate the axion mass.

\subsection{Misalignment mechanism}

As usual in standard cosmology, the universe will be described by a flat FRW metric \cite{turner:kolb:cosmology}, with cosmological parameters given by the concordance of the best available data (we take WMAP5+BAO+SN \cite{komatsu:wmap5:cosmological:interpretation}). For the temperature regions of interest we can restrict ourselves to radiation and axions, in which case Einstein's equations are given by
\begin{eqnarray}
&H^2 = \frac{1}{3M_p^2}\left(\frac{\pi^2}{30}g_{*,R}T^4+f_a^2 \left(\frac{1}{2}\dot{\theta_a}^2+m_a^2(T)(1-\cos\theta_a)\right)\right)& \,,\\
&\ddot{\theta} + 3 H \dot{\theta_a} + m_a^2(T) \sin \theta_a = 0 \label{eq:axion:evolution}& \,,
\end{eqnarray}
where $M_P^2$ is the reduced Planck mass. Note that the effective axion potential has been shifted so that non-perturbative effects do not lead to a non-vanishing vacuum energy\footnote{Note that there exist theories that combine another axion-like field to entangle the dark matter and the dark energy sector \cite{nilles:quintaxion,kim:axion:quintessential}.}.

\begin{figure}[tbp]
\begin{center}
 \includegraphics[width=\figwidth,clip=true,trim=0mm 0mm 15mm 10mm]{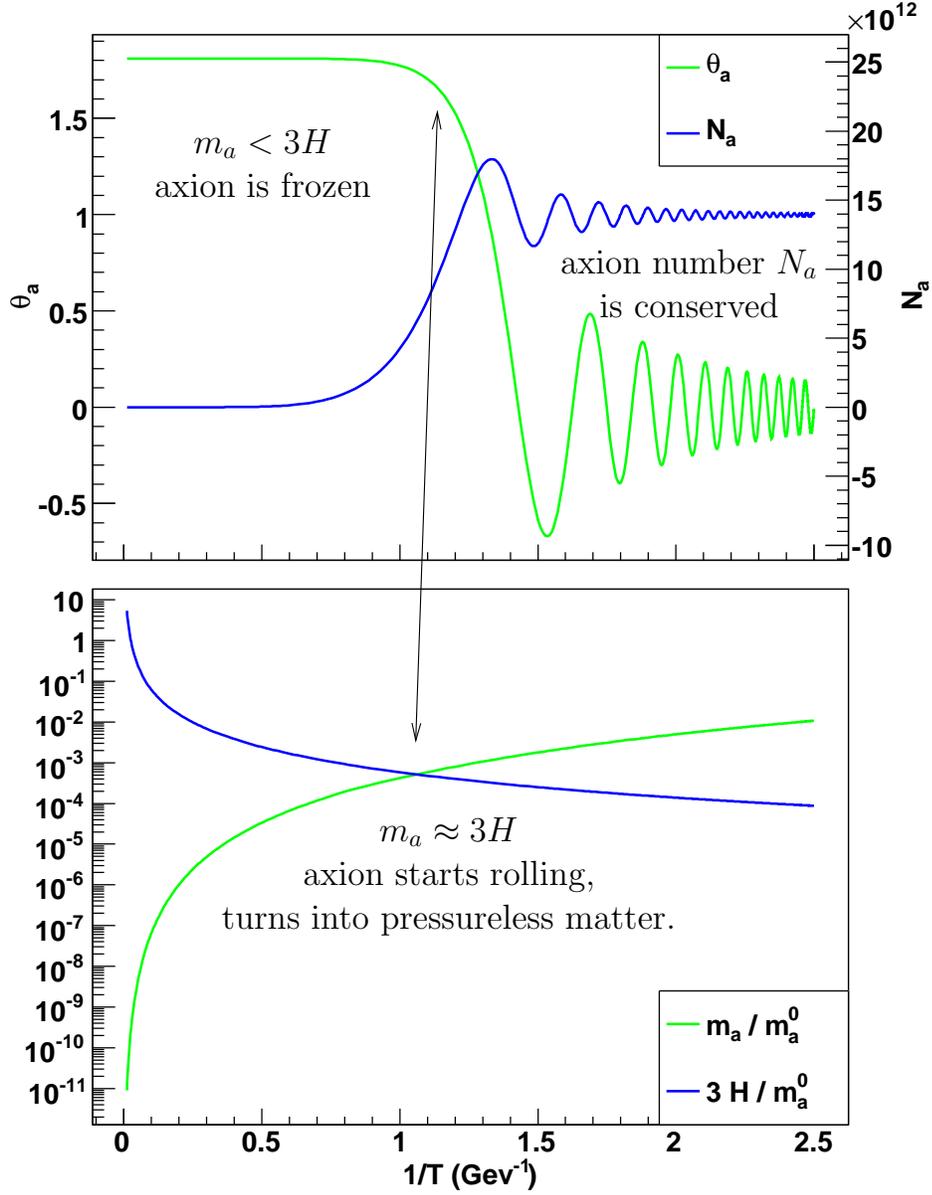}
\end{center}
\caption{As long as the axion Compton wavelength is well outside the horizon, the axion zero mode is frozen; this corresponds to the late-time solution of (\ref{eq:axion:evolution}) with $m_a$ neglected. The axion starts to feel the pull of its mass at $m_a \approx 3H$, and evolves to its minimum at $\theta_a=0$, i.e.\ the PQ mechanism to solve the strong CP problem. After a few oscillations the axion number per comoving volume stays constant as long as the axion mass and the scale factor change slowly (adiabatic approximation). This is then used to extrapolate the result to today.}\label{fig:dynamics}
\end{figure}

The dynamics of the axion evolution consists of three qualitatively different stages: First, as long as its Compton wavelength is above the Hubble scale, the axion is effectively massless; the Hubble friction enforces a constant axion field in this case. Secondly, once the axion mass becomes comparable to the Hubble scale, at a time when $m_a \approx 3H$ holds, the axion feels the pull of its mass and starts to roll towards its minimum at $\theta_a=0$. Finally, after a few oscillations the axion evolution is indistinguishable from pressureless matter and the axion number per comoving volume is conserved. These three regimes are illustrated clearly for an explicit numerical solution in \reffig{fig:dynamics}. 

The physics underlying the misalignment mechanism is based on the fact that the energy redshifts with time, and that the Hubble dilution starts once the oscillations in the axion zero mode begin.  Consequently, the total Hubble redshift increases with $m_a$. This leads to the a priori counterintuitive behaviour that light axions, i.e.\ high $f_a$, contribute more to the energy balance than heavy axions.

\begin{figure}[tbp]
\begin{center}
 \includegraphics[width=\figwidth,clip=true,trim=0mm 0mm 15mm 10mm]{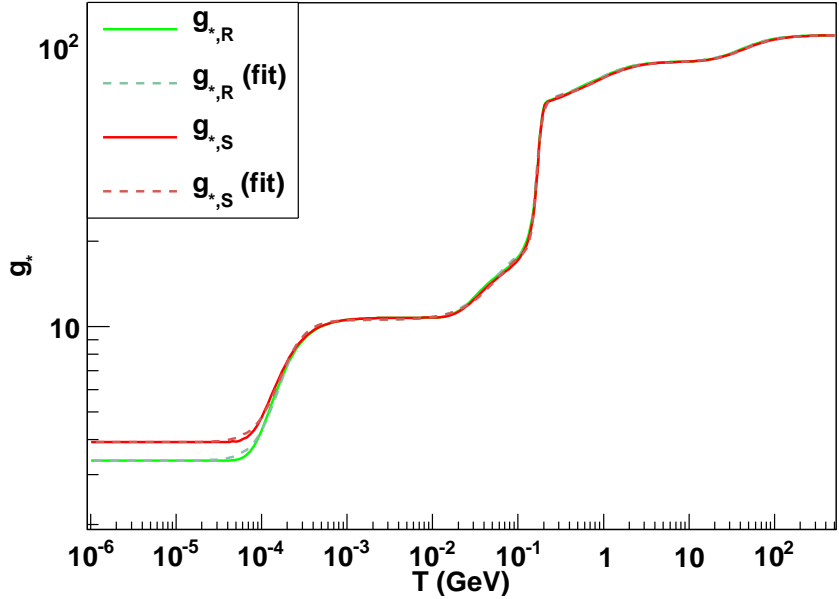}
\end{center}
\caption{The effective degrees of freedom $g_{*,R}$ and $g_{*,S}$ are given for the temperature range up to $T\approx 100\units{GeV}$. The decoupling of the neutrinos is included and manifests itself in the differences between $g_{*,R}$ and $g_{*,S}$ after $e^\pm$ annihilation, when $T_\nu \neq T_\gamma$. We followed closely \cite{coleman:roos:geff}, but included some minor changes to take into account a better understanding of the QCD phase transition from recent lattice studies (see main text). We have determined fits by using a sequence of smoothed step functions, see appendix \ref{app:geff:fit}. As seen from the graph, the fits are good, generally with an accuracy below $1\%$.}\label{fig:geff}
\end{figure}

Analytical progress can be made by noting that
\begin{equation}
\frac{\dot{\rho}_a}{f_a^2} = 2m_a\dot{m}_a(1-\cos\theta_a)-3H\dot{\theta_a}^2 \,,
\end{equation}
where we have made use of the equation of motion for the axion field (\ref{eq:axion:evolution}). We assume that over one oscillation $a$, $H$ and $m_a$ do not change much, i.e.\ adiabatic evolution. Furthermore, we consider times when the axion field has been Hubble redshifted for long enough so that anharmonic effects are negligible, in which case the axion behaves as a damped harmonic oscillator with constant coefficients. For such a system $\overline{\dot{\theta}^2_a}=m^2_a \overline{\theta^2_a}$, where the expectation value is an average over one oscillation. This leads to
\begin{equation}
\frac{\overline{\rho_a}a^3}{m_a} = \mathrm{const} \,. \label{eq:axion:number:conservation}
\end{equation}
\eq{\ref{eq:axion:number:conservation}} allows us to easily compute the energy in the axion field today
\begin{equation}
\rho_a(\mathrm{today}) = \rho_a(T)\frac{m_a(\mathrm{today})}{m_a(T)}\frac{s(\mathrm{today})}{s(T)}\, ,\label{eq:axion:energy:density:today}
\end{equation}
where $T$ is the temperature when we reached the asymptotic behaviour predicted by (\ref{eq:axion:number:conservation}), $s=\frac{2\pi^2}{45}g_{*,S}T^3$ is the entropy density and $g_{*,S}$ is the effective number of degrees of freedom shown in Fig.~\ref{fig:geff}; we have neglected the overbar for simplicity. This can then be compared to the present critical energy density, which constrains the axion mass. Note that apart from the time average, a spatial averaging is implicitly understood. In the formulas that follow the coarse-grained $\theta_a$ now stands for the effective axion angle 
\begin{equation}
\theta_a^\mathrm{eff}=\sqrt{\langle \theta_a^2 \rangle} = \sqrt{\langle \theta_a \rangle^2 + \sigma_{\theta_a}^2}\,,
\end{equation}
where $\sigma_{\theta_a}$ represents contributions from quantum fluctuations and classical inhomogeneities. (This notation for $\theta_a$ is used except where explicitly stated otherwise or in expectation values.).

As alluded to in the introduction, the axions produced from the heat bath have been ruled out as the dominant cold dark matter contribution \cite{hannestad:mirizzi:raffelt:thermal:axion}. The remaining window for larger values of $f_a$, bounded from below by astrophysical considerations, implies tiny couplings of axions to ordinary matter so that the axions will stay out of equilibrium with respect to the cosmic plasma; anharmonic effects that couple higher modes can also be ignored by the same reasoning. In addition, axions have very small velocity dispersions\footnote{The zero-mode has, by definition, none. Non-zero-mode contributions, due to inhomogeneities in the axion field set up during the PQ phase transition and due to string decay, can be shown to be small (see, for example, Sikivie's review article in \cite{kuster:raffelt:beltran:axions}).}. Therefore, axions provide a generic cold dark matter candidate, which is constrained by observation to be $\Omega_c\le 0.23$ \cite{komatsu:wmap5:cosmological:interpretation}.

We will follow the early paper \cite{turner:axion:cosmology} and the recent update \cite{bae:huh:kim:axion} to derive an analytic formula for the axion abundance which we then compare to a full numerical solution (using the same updated parameters).  We will use the simple power-law approximation (\ref{eq:mass:dga}) for the axion mass from  the dilute instanton gas; it reaches the $T=0$ mass at $T\approx 100 \units{MeV}$, converges to the full IILM mass at $T\approx 400 \units{MeV}$ and overestimates it slightly for higher temperatures, as 
can be seen in \reffig{fig:masses}. Note that the coefficient $\alpha_a \propto m_u m_d m_s$. The axion starts rolling at
\begin{eqnarray}
 T_a^{n+4} &\approx& 0.04 \frac{\Lambda^{n+4} \alpha_a m^2_{Pl}}{g_{*,R} f^2_a}\,, \quad T \gtrsim 103\units{MeV}\label{eq:T:oscillation:high}\\
 T_a^4~~ &\approx& ~0.04 \frac{\chi m^2_{Pl}}{g_{*,R} f^2_a} \,,~\qquad T \lesssim 103\units{MeV}.\label{eq:T:oscillation:low}
\end{eqnarray}
Assuming that the number of axions per comoving volume is conserved at $T_a$, we can use (\ref{eq:axion:energy:density:today}) to get
\begin{equation}
 \rho_a=\frac{45 s_0}{4\pi^2} \left\{
 \begin{array}{c}
  (0.2\,m_{Pl})^{-\frac{6+n}{4+n}} g_{*,R}^{-\frac{1+n/2}{4+n}} \chi^{\frac{1}{2}}  \Lambda^{-1} \alpha_a^{-\frac{1}{4+n}} f_a^{\frac{6+n}{4+n}} \theta_a^2\\
  (0.2\,m_{Pl})^{-\frac{3}{2}} g_{*,R}^{-\frac{1}{4}} \chi^{\frac{1}{4}} f_a^{\frac{3}{2}}  \theta_a^2
 \end{array}
 \right. .\label{eq:axion:energy:density}
\end{equation}
Note that the axion abundance is fairly insensitive to changes in $\alpha_a$ due to its small exponent. In particular, the large uncertainties in the quark masses have only a minor effect on the final results. We therefore expect this estimate of the energy density $\rho_a$ to differ only slightly relative to the results of \cite{bae:huh:kim:axion}, say. The additional $T$-dependence due to $g_{*,R}$ and $g_{*,S}$ will also be weak, again due to rather small exponents. Using $\Lambda=0.4\units{GeV}$, $\chi=(78.2\units{MeV})^4$ and $T_\gamma = 2.725 \units{K}$ we get 
\begin{equation}
 T \approx \left\{
 \begin{array}{c}
  1.46 \units{GeV}\, g_{*,R}^{-0.094} \left(\frac{10^{12}\units{GeV}}{f_a}\right)^{0.19} \\
  122 \units{GeV}\, g_{*,R}^{-0.25} \left(\frac{10^{12}\units{GeV}}{f_a}\right)^{0.5}
 \end{array}
 \right. ,
\end{equation}
and
\begin{equation}
 \Omega_a h^2 \approx \left\{
 \begin{array}{c}
  0.54\, g_{*,R}^{-0.41} \theta_a^2 \left(\frac{f_a}{10^{12}\units{GeV}}\right)^{1.19} \\
  0.0064\, g_{*,R}^{-0.25} \theta_a^2 \left(\frac{f_a}{10^{12}\units{GeV}}\right)^{1.5}
 \end{array}
 \right. .
\end{equation}

These analytic results can be improved by taking anharmonic effects into account \cite{bae:huh:kim:axion}. The upshot is that we can incorporate these through the substitution $\theta^2_a \to \theta^2_a f(\theta_a)$, where
\begin{equation}
 f(\theta_a) = \frac{4\sqrt{2}}{\pi\theta_a^2} \int_0^{\theta_a}d\theta\, \sqrt{\cos\theta-\cos\theta_a} \,,
\end{equation}
with $f(\theta_a) \to 1$ as $\theta_a \to 0$; it decreases monotonically to $f(\pi) \approx 0.516$. In this regime, $\theta_a \to \pi$, the adiabatic condition breaks down because the axion behaves like an inflaton, and the scale factor $a$ no longer varies slowly. However, analytic progress can be made in this limit if one takes into account that the above quantity needs to be combined with an estimate for the time when the axion actually starts rolling (and the inflationary phase is over). This occurs no longer at $m_a \approx 3H$ \cite{lyth:axion:inflation:fluctuations:1,strobl:weiler:axion:anharmonic,visinelli:gondolo:axion}. In the full numerical set-up, we always propagate the solution into the harmonic regime before we use (\ref{eq:axion:energy:density:today}) to extrapolate to today, and so the correction factor $f(\theta_a)$ is not required.

To take the temperature dependence of $g_*(T)$ into account, we must solve the implicit equations (\ref{eq:T:oscillation:high}) and (\ref{eq:T:oscillation:low}). We find that neglect of the temperature dependence of $g_*$ leads to an error of $10\%$ in the analytic computation. On top of that the analytic computation assumes that the adiabatic regime is reached when $m_a=3H$; this is a further source of error relative to the full numerical result.

\begin{figure}[tbp]
\begin{center}
\includegraphics[width=\figwidth,clip=true,trim=0mm 0mm 15mm 10mm]{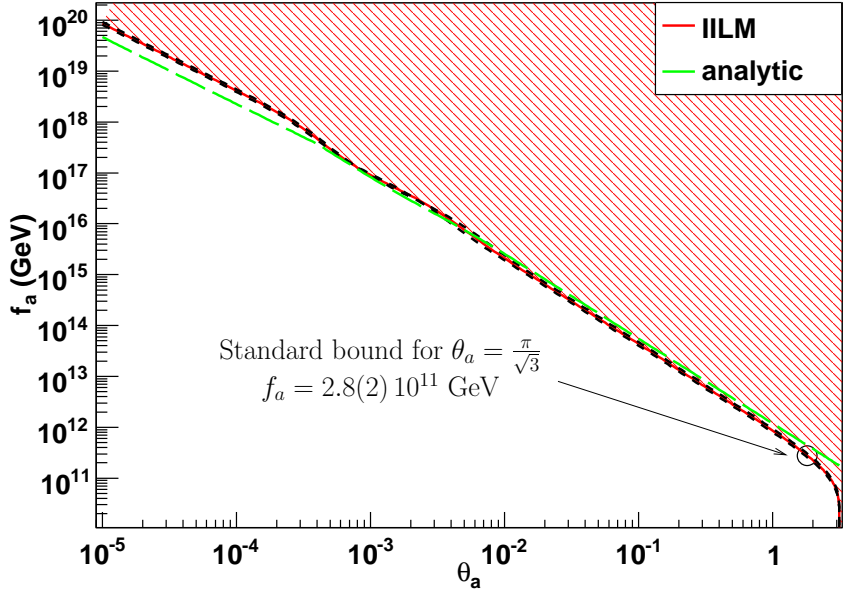}
\caption{The anthropic axion is defined through its relation between $f_a$ and $\theta_a$ given a fixed $\Omega_a$. We display here the result for the case that axions form the dominant dark matter component of the universe, i.e.\ $\Omega_a=0.23$. The short-dashed lines correspond to the uncertainties in $f_a(\theta_a)$ from the systematic errors in $\Lambda$, which is the dominant source of uncertainties. We also include the result of the analytic computation (\ref{eq:axion:energy:density}). Over many orders of magnitude the agreement is very good. More pronounced differences only show up at $\theta_a \to 0$ and $\theta_a \to \pi$. The latter is due to the fact that the adiabatic condition is not fulfilled because the potential becomes very flat and acts like a source of inflation, i.e.\ a rapidly changing scale factor. The differences between the analytic and numerical data at small $\theta_a$ follow from the different functional form of the axion masses: in the former case the axion mass is constant whereas in the latter it is slightly rising with temperature (see main text).}
\label{fig:ThetaFa}
\end{center}
\end{figure}

Imposing the dark matter constraint $\Omega_\mathrm{c} \approx 0.23$, we get a relation between the two parameters $f_a$ an $\theta_a$. The analytic approximation compares well with the full numerical result, except for very large or very small initial misalignment angles, see \reffig{fig:ThetaFa}. At large angles the adiabatic condition breaks down. At small angles the dilute gas approximation (DGA) to the axion mass is constant whereas the full IILM mass decreases slowly towards $m_a(T=0)$, i.e.\ the IILM dynamics starts rolling slightly earlier so that the Hubble dilution acts for longer which in turn leads to a smaller $\Omega_a$ at a given $f_a$; equivalently, to reach $\Omega_c$ a slightly larger $f_a$ is needed, as depicted in \reffig{fig:ThetaFa}.

If the PQ symmetry breaks spontaneously after inflation, the correlated domains with a given misalignment angle are not  stretched to macroscopic sizes and a strong spatial dependence remains. In ref.~\cite{turner:axion:cosmology}, these fluctuations are averaged in the observable universe to find the root-mean-square fluctuations $\theta_a=\frac{\pi}{\sqrt{3}}$. This was then taken to be the initial condition for an estimate of the zero momentum mode axions. While this approach to axion production in the thermal scenario is flawed (as we shall discuss), it provides a useful benchmark with which to compare competing estimates. Adopting these initial conditions, from the full numerical results we obtain the important bound
\begin{equation}
 f_a\;\le\;2.8(\pm2)\times10^{11}\units{GeV} \quad \hbox{or} \quad  m_a\;\ge\;21(\pm2) \units{\mu eV} \,.\label{eq:misalignment:bound}
\end{equation}
This canonical result can be compared directly with our analytic modelling, as well as other estimates in the literature, to check accuracy.

\begin{figure}[tbp]
\begin{center}
\includegraphics[width=\figwidth,clip=true,trim=0mm 0mm 15mm 10mm]{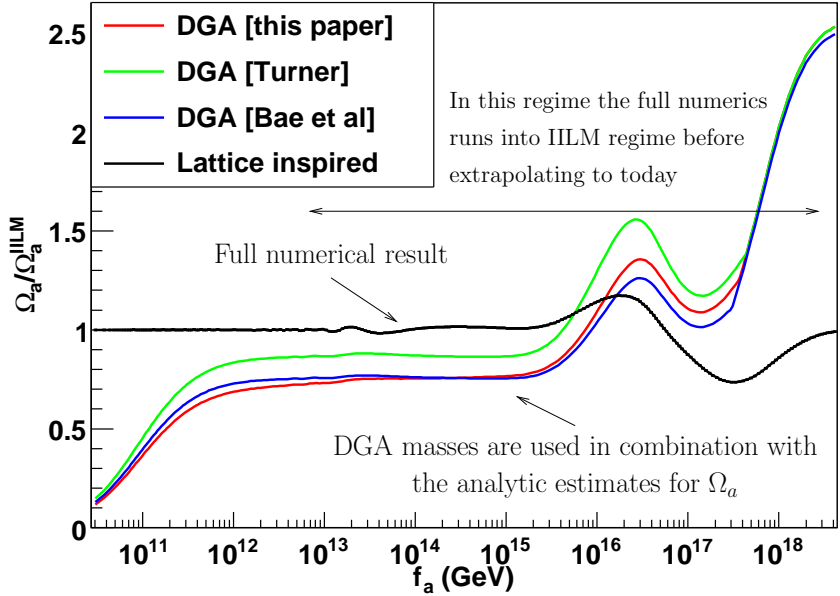}
\caption{To analyse the differences between the different determinations of the axion abundance, we compute $\Omega_a$ on a fixed set of $\{f_a,\theta_a\}$ derived from the full numerical determination with the IILM axion mass, given that $\Omega_a=\Omega_c$. We see that the analytic results are off by more than a factor of $2$ in the regimes of small and large axion angles; for the case when inflation has not operated after PQ symmetry breaking, i.e.\ for $f_a<2.8\times10^{11}\units{GeV}$, we can see that the numerical and analytic results are different by a factor of $3$ (and more if the axion is not the dominant dark matter component). For the main part of parameter space the discrepancy is smaller but systematically an underestimate. The full numerical data using the lattice inspired axion mass are very similar for most $f_a$ (by construction), and differ only for large $f_a$, as might have been expected. It is noteworthy that the numerical determination runs into the proper IILM axion mass already for rather low $f_a$'s before it extrapolates to today.}
\label{fig:OmegaFa}
\end{center}
\end{figure}

In \reffig{fig:OmegaFa}, the errors of the analytic results with the DGA axion masses are compared to the numerical result across the full range of $f_a$. The numerical solution has $f_a$ and $\theta _a$ chosen such that $\Omega_a= 0.23$. Except at large and small $f_a$, the differences with the analytic models using the same parameters are of order $O(1)$.  It is clear in this regime that the relative abundance is not strongly dependent on the exact shape of the axion mass. We also compared the full numerical results for the DGA masses and found again that differences are only pronounced in the region of the QCD phase transition where the different mass ans\"atze differ considerably. We checked against the full numerical solution for the lattice-inspired mass function too, see \reffig{fig:masses}, and found less than $5\%$ variations for  $f_a< 10^{15} \units{GeV}$. 

While revealing the regimes in which analytic estimates go astray, for the most part the results are in 
good agreement with this simple DGA treatment. It is encouraging to note that estimates of the axion density are 
relatively insensitive to the detailed nature of the axion mass function, affirming the validity of the previous
literature. However, this does not mean that a simple treatment will automatically give agreement with 
(\ref{eq:misalignment:bound}); this requires appropriate normalisation, updated parameter choices and a careful 
treatment of $g_*$, as attested by the significant differences between quoted bounds.

\subsection{Anthropic window, isocurvature and fine-tuning bounds}

The discussion so far has been purely classical. In order to discuss the anthropic window, where we fine-tune $\theta_a$ at large and small $f_a$, we need to take into account the quantum mechanical fluctuations of the axion field induced by inflation. Like any other massless field, the axion receives quantum mechanical fluctuations during the quasi de~Sitter evolution of the early universe, i.e.\ $\sigma_{\phi} = \frac{H_I}{2\pi}$ or in terms of the axion angle
\begin{equation}
\sigma_{\theta_a} = \frac{H_I}{2\pi f_a} \,.
\end{equation}
Apart from the spatial averages, these quantum mechanical effects need also be taken into account in the coarse grained equations; they are dominant for the anthropic scenario where inflation has smoothed out the `classical' inhomogeneities, and makes sense only if the PQ symmetry is broken before the end of inflation. The fluctuations in the massless axion field then lead to isocurvature perturbations in the cosmic microwave background radiation (CMBR). The ensuing constraints apply only if the PQ symmetry has not been restored after inflation; this could happen if the quantum mechanical fluctuations or the reheat temperature are too large, i.e.\ if $f_a<\max\left[\frac{H_I}{2\pi},T_\mathrm{RH}\right]$.

As mentioned previously, the fluctuations lead to an effective axion angle, $\theta_a^\mathrm{eff}=\sqrt{\langle \theta_a^2 \rangle}=\sqrt{\langle \theta_a \rangle^2 + \sigma^2_{\theta_a}}$, and will be used in the coarse grained evolution equations. It follows that the effective angle is bounded by the root-mean-square fluctuations
\begin{equation}
 \theta_a^2 \ge \sigma_{\theta_a}^2 = \frac{H^2_I}{(2\pi f_a)^2} \,.\label{eq:isocurvature:constraint:1}
\end{equation}

We will follow refs~\cite{burns:axion:isocurvature,beltran:garcia-bellido:lesgourges:isocurvature,hertzberg:tegmark:wilczek:axion,hamann:hannestad:raffelt:wong:isocurvature,visinelli:gondolo:axion} to put constraints on the PQ scale $f_a$ and on the inflationary scale $H_I$. By definition, isocurvature perturbations do not change the energy density, i.e.\ $\delta \rho = 0$. For a massless field such as the axion, fluctuations do not change the potential energy density. In addition we have assumed that the quantum mechanical fluctuations are small, i.e.\ $f_a>H_I$ so that the PQ symmetry is not restored, and the gradient energy density is negligible compared to the inflationary energy scale, for details see ref.~\cite{beltran:garcia-bellido:lesgourges:isocurvature}. Thus, inflationary axion fluctuations are indeed isocurvature which leads to
\begin{equation}
 \delta \rho =0= \delta \rho_a + \sum_{i\neq a}\delta \rho_i + \delta \rho_r\,.
\end{equation}
Assuming that all other fields have adiabatic perturbations, i.e.\ $\frac{\delta \rho_i}{\rho_i} = \frac{3}{4} \frac{\delta \rho_r}{\rho_r}$, we get a relation between the temperature fluctuation, i.e.\ $\frac{\delta \rho_r}{\rho_r}=4\frac{\delta T}{T}$, and the axion fluctuation
\begin{equation}
 \frac{\delta T}{T}=-\frac{\rho_a}{3\sum_{i\neg a}\rho_i + 4 \rho_r} \frac{\delta n_a}{n_a}\,.\label{eq:temperature:fluctuation}
\end{equation}

It is customary to define the entropy perturbation
\begin{equation}
 S_i \equiv \frac{\delta(n_i/s))}{n_i/s}=\frac{\delta n_i}{n_i}-3\frac{\delta T}{T}\,,
\end{equation}
where $s\propto T^3$ is the entropy density and $n_i$ the number density of particle species $i$. It is clear that for adiabatic perturbations $S_i=0$ by definition. At the time the relevant perturbations are set, the universe is radiation dominated, and it follows from (\ref{eq:temperature:fluctuation}) that the radiation perturbations are negligible with respect to axion fluctuations. The initial condition is thus given by $S_a = \frac{\delta n_a}{n_a}$ and once this mode leaves the horizon it remains constant.

The relevant scales cross back inside the horizon during matter domination, i.e.\ $\rho_r \ll \rho_a$; it then follows from (\ref{eq:temperature:fluctuation}) that
\begin{equation}
 \left(\frac{\delta T}{T}\right)_\mathrm{iso} = -\frac{6}{15}\frac{\Omega_a}{\Omega_m} S_a\,,
\end{equation}
where we have added the Sachs-Wolfe contribution and $\Omega_m$ is the total matter abundance.
The fraction of isocurvature to total temperature fluctuations has been constrained to \cite{komatsu:wmap5:cosmological:interpretation}
\begin{equation}
 \alpha_a \equiv \frac{\langle(\delta T/T)^2_\mathrm{iso}\rangle}{\langle(\delta T/T)^2_\mathrm{tot}\rangle} < 0.072\,,
\end{equation}
at $95\%$ confidence level, and $(\delta T/T)^\mathrm{rms}_\mathrm{tot} \approx 1.1 \times 10^{-5}$, where the sum is taken over the first few low-lying multipoles.

For the large $f_a$ we are interested in, the axion dependence on the energy density is well approximated by $\rho_a \propto \theta^2_a$, where in this case $\theta_a$ is not the effective axion angle. Assuming a Gaussian distribution for the axion angle perturbation, we find that
\begin{equation}
 \langle S^2_a \rangle = \left\langle \left(\frac{\theta_a^2-\langle \theta_a^2 \rangle}{\langle \theta_a^2 \rangle}\right)^2 \right\rangle = \frac{2\sigma_{\theta_a}^2(2\theta_a^2-\sigma_{\theta_a}^2)}{\theta_a^4}\,.\label{eq:entropy:perturbation:harmonic}
\end{equation}
From this it follows that the isocurvature fraction is given by
\begin{equation}
\alpha_a = \frac{4}{25} \frac{\Omega_a^2/\Omega_m^2}{\langle(\delta T/T)^2_\mathrm{tot}\rangle} \frac{2\sigma_{\theta_a}^2(2\theta_a^2-\sigma_{\theta_a}^2)}{\theta_a^4}\,,
\end{equation}
which we can rewrite into a constraint equation for $H_I$:
\begin{equation}
 \left(\frac{H_I}{2\pi}\right)^4 - 2 (\theta_a f_a)^2 \left(\frac{H_I}{2\pi}\right)^2 + \frac{\tilde{\alpha}_a}{2\Omega_a^2} (\theta_a f_a)^4 > 0\,.
\end{equation}
All explicit numerical factors have been absorbed into 
\begin{equation}
 \tilde{\alpha}_a = 0.072 \frac{25}{4} \langle(\delta T/T)^2_\mathrm{tot}\rangle \Omega_m^2\approx4\times10^{-12}\ll 1\,.
\end{equation}
Of the two possible solutions, the larger one is in conflict with (\ref{eq:isocurvature:constraint:1}), so that the isocurvature constraint becomes
\begin{equation}
 H_I < \frac{\sqrt{\tilde{\alpha}_a}\pi}{\Omega_a} \theta_a f_a < \frac{\sqrt{\tilde{\alpha}_a}\pi}{\Omega_a} \theta_a(f_a) f_a \approx \frac{6.3\,10^{-6}}{\Omega_a}\theta_a(f_a) f_a\,, \label{eq:axion:isocurvature:harmonic}
\end{equation}
where $\theta_a(f_a)$ follows from \reffig{fig:ThetaFa}. Note that $\theta_a/\Omega_a \propto \Omega_a^{-\frac{1}{2}}$, so that for fixed $f_a$ the bound becomes weaker if the axions make up only a fraction of the dark matter content of the universe. 

The dependence of the isocurvature bound on the axion mass is encoded in the function $\theta_a(f_a)$ for fixed $\Omega_a$. From \reffig{fig:OmegaFa} we can already anticipate that the exact numerical result will not depend strongly on the axion mass. We checked explicitly that the dependence on the masses is rather small,
\begin{equation*}
 \frac{2}{3} < \frac{\theta_a f_a}{(\theta_a f_a)^\mathrm{IILM}} < \frac{4}{3}\,.
\end{equation*}
In fact, these largest discrepancies occur around $\theta_a \approx O(10^{-3})$. For smaller angles the results converge, whereas for larger angles the differences are on the $10\%$ level, driven by the different exponents with which the masses decay. 

Actually, (\ref{eq:axion:isocurvature:harmonic}) is an implicit equation for $H_I$ since the effective axion angle depends on it. However, in this regime the difference between $\theta_a$ and $\langle \theta_a \rangle$ is negligible. To see this, we rewrite (\ref{eq:axion:isocurvature:harmonic}) in terms of $\langle \theta_a \rangle$, and get
\begin{equation}
 H_I < \frac{\sqrt{\tilde{\alpha}_a}\pi}{\Omega_a} \langle \theta_a \rangle f_a \left(1-\frac{\tilde{\alpha}_a}{4\Omega_a^2} \right)^{-\frac{1}{2}} \approx \frac{\sqrt{\tilde{\alpha}_a}\pi}{\Omega_a} \langle \theta_a \rangle f_a\,.
\end{equation}
This, in turn, can be used to estimate that $\sigma_a \ll \langle \theta_a \rangle$. In the anthropic regime the dependence of $\theta_a$ on $H_I$ is totally negligible, and we can fine-tune the effective axion angle to $\theta_a \approx \langle \theta_a \rangle (1+10^{-10})$.

If we want to extend the analysis to all of the anthropic window, i.e.\ $\theta_a=O(1)$, we need to take into account anharmonic effects. We are greatly helped by the fact that $\sigma_a \ll \langle \theta_a \rangle$: it allows us to Taylor expand the true axion potential around $\langle \theta_a \rangle$ and we find that
\begin{equation}
\langle S^2_a \rangle = \frac{\sin^2\langle\theta_a\rangle (\sigma_{\theta_a}^2-\sigma_{\theta_a}^4) + \frac{1}{2} \cos^2\langle\theta_a\rangle \sigma_{\theta_a}^4}{(1-\cos\langle\theta_a\rangle)^2 + \frac{1}{4}\cos^2\langle\theta_a\rangle \sigma_{\theta_a}^4 + (1-\cos\langle\theta_a\rangle)\cos\langle\theta_a\rangle \sigma_{\theta_a}^2(1-\frac{1}{4}\sigma_{\theta_a}^2)}\,,\label{eq:entropy:perturbation:anharmonic}
\end{equation}
where we expanded to fourth order, and we assume again that $\delta \theta_a$ is Gaussian with mean zero. For small $ \sigma_{\theta_a} \ll \langle \theta_a \rangle \ll 1$ this goes over into (\ref{eq:entropy:perturbation:harmonic}). For large misalignment angles the behaviour becomes rather different, and for $\langle \theta_a \rangle \to \pi$ we see that the entropy perturbation tends to $\langle S^2_a \rangle \to \frac{1}{8}\sigma_{\theta_a}^4$ in contrast to the $\langle S^2_a \rangle \to \frac{4}{\pi^2}\sigma_{\theta_a}^2$ for the harmonic case. Since the perturbation has to be small, the solution for $\sigma_{\theta_a}$ in the anharmonic case will be much larger and the constraint on $H_I$ consequently much weaker. The solution satisfies again $\sigma_a \ll \langle \theta_a \rangle$, and the approach is self-consistent.

\begin{figure}[tbp]
\begin{center}
\includegraphics[width=\figwidth,clip=true,trim=0mm 0mm 15mm 10mm]{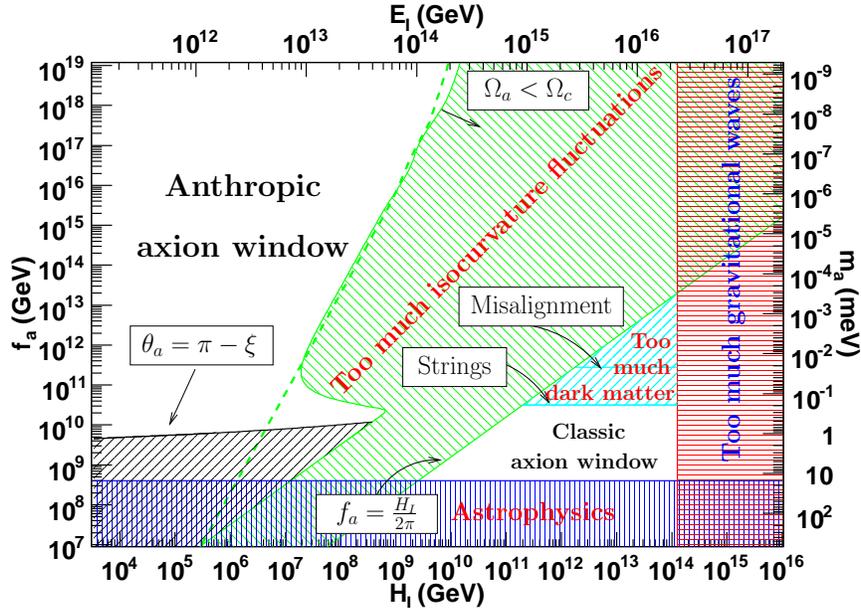}
\caption{The allowed parameter space in the $H_I$--$f_a$ plane is plotted in white; the inflationary energy scale is defined by $E_I \equiv (H_I m_{Pl}/\sqrt{8\pi/3})^{1/2}$. The green curve in the upper left corner follows from the isocurvature constraint (\ref{eq:entropy:perturbation:anharmonic}), when the axion is the dominant dark matter candidate; the dashed line corresponds to the (semi-)analytic computation (based on (\ref{eq:T:oscillation:high})-(\ref{eq:axion:energy:density}), taking fully into account the temperature dependence of $g_*$) together with (\ref{eq:axion:isocurvature:harmonic}), i.e.\ the constraint for a harmonic potential. If axions provide only a fraction of the dark matter content of the universe, the bound becomes weaker. For lower $f_a$, i.e.\ larger $\theta_a$, the anharmonic effects become important and the bound on $H_I$ weakens because anharmonic effects lead to smaller perturbations. For $\theta_a \to \pi$, the dependence of $H_I$ on $\theta_a$ can no longer be neglected and leads to the black curve. The lower green curve gives the lower bound for isocurvature production (very inefficient reheating is assumed \cite{hertzberg:tegmark:wilczek:axion}); beneath this curve, the axion angle is spatially varying (with root-mean-square fluctuation $\theta_a={\pi}/{\sqrt{3}}$). The cyan wedge is excluded as it would lead to too much dark matter from axion string radiation. The bound from the non-detection of gravitational waves, i.e.\ $r<0.22$ \cite{komatsu:wmap5:cosmological:interpretation}, leads to the upper bound on the inflationary scale $H_I<1.26\times10^{14}\units{GeV}$. Finally, $f_a$ is bounded from below by astrophysical considerations, i.e.\ axion emission from stars; we use $f_a>4\times10^8 \units{GeV}$, see \cite{kuster:raffelt:beltran:axions} chapter 3.}
\label{fig:parameter:space}
\end{center}
\end{figure}

Even though the fluctuation is very small, the regime $\theta_a \to \pi$ can only be achieved if $H_I$ is further constrained, as we will now show. Fine-tuning in the effective axion angle is hindered at large $H_I$ by quantum fluctuations. Assuming that $\langle \theta_a \rangle = \pi$, we obtain the following bound on $H_I$
\begin{equation}
 H_I < \sqrt{8\pi\xi} \pi f_a\,,
\end{equation}
where $\xi$ is defined by $\xi\equiv\pi-\theta_a$ and encodes the degree of fine-tuning. It is clear then that for axions on the top of the potential, this bound supersedes the isocurvature bound. This leads to a further restriction on the  allowed parameter space shown in \reffig{fig:parameter:space} (bottom left). This cut-off provides an interesting boundary in parameter space suggesting that inflationary axions have a maximum possible mass of approximately $m_a\lesssim 1\units{meV}$. Inflationary scenarios usually consider small axion masses in the range $m_a \lesssim 10\units{\mu eV}$, but we see here with anharmonic tuning that it is possible to have $\Omega_a = 0.23$ with masses at and above that of the thermal dark matter axion (see below). Naturalness arguments tend to disfavour inflation models with a low energy scale, but we note that the weakening of the isocurvature constraint at $\theta_a \approx \pi$ opens up the inflationary window by several orders of magnitude relative to naive expectations. It is possible to have these heavier axions with an inflation scale approaching $H\approx 10^9 \units{GeV}$. The fact that two alternative scenarios -- thermal and inflationary -- can produce axions with masses $m_a = 0.1$--$1\units{meV}$ might strengthen the case for experimental searches in this parameter regime.

Note that although the parameter space is really three-dimensional, i.e.\ in $\langle \theta_a \rangle$, $f_a$ and $H_I$, we only plot the $H_I$--$f_a$ plane as these are the fundamental parameters, whereas $\langle \theta_a \rangle$ is naturally seen as an `environmental' feature of our location in the universe after inflation. Recall also that without inflation the causal patches with different $\langle \theta_a \rangle$ stay microscopic all the way to today with $\theta_a \to \frac{\pi}{\sqrt{3}}$, so again the axion angle is not required as an extra parameter. The anthropic window bounds can be weakened by choosing $\theta_a<\theta_a(f_a)$, however, if the axion is to be the dominant dark matter candidate then the parameter space is truly 2-dimensional and collapses to the $H_I$--$f_a$ plane.

We have mentioned previously, that the natural axion angle to be used in the evolution equations is the effective axion angle that takes into account root-mean-square fluctuations. A priori this entangles $H_I$ and $\langle \theta_a \rangle$. However, we have seen that for a large part of the anthropic window $\theta_a$ does not really depend on $H_I$. On the upper green curve in \reffig{fig:parameter:space}, that is if the axion is to be the dominant dark matter candidate, then $f_a$ and $\langle \theta_a \rangle$ are in the one-to-one correspondence defined through \reffig{fig:ThetaFa}. Again, on the black curve in \reffig{fig:parameter:space}, $\langle \theta_a \rangle$ is very close to $\pi$ and $\theta_a$ depends solely on $H_I$ for axions that contribute $\Omega_a=\Omega_c$.

\begin{figure}[tbp]
\begin{center}
\includegraphics[width=0.6\figwidth]{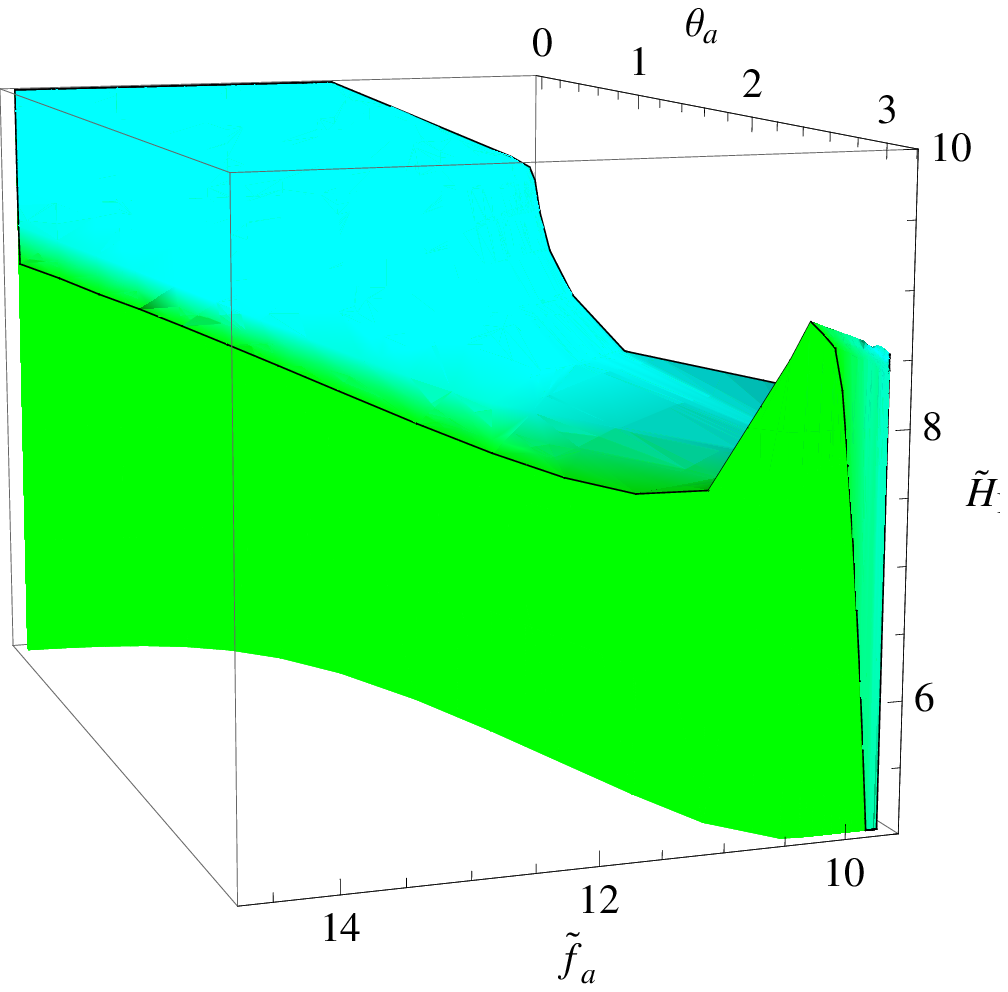}
\includegraphics[width=0.6\figwidth]{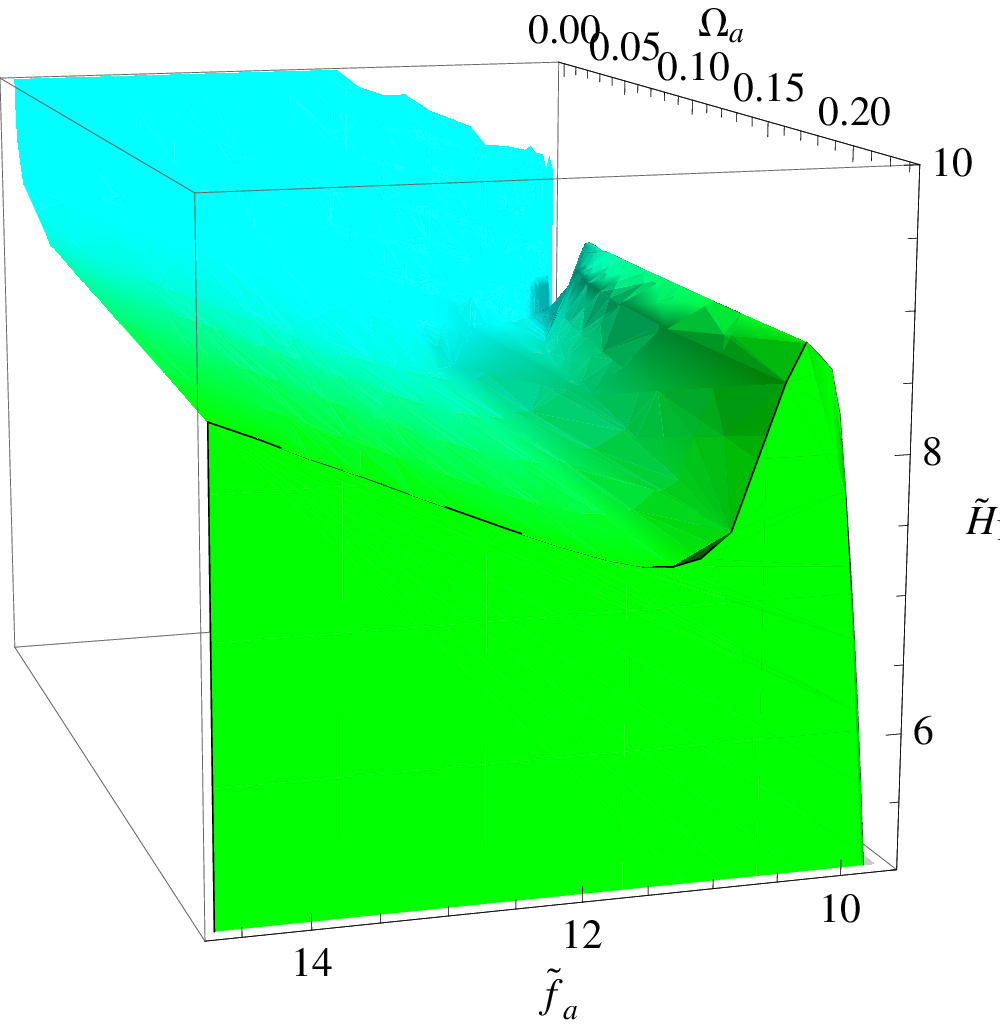}
\caption{The allowed parameter space in the anthropic window. The dark matter axion lives on the face pointing towards the $H_I$--$f_a$ plane. Although not as clearly visible as for the right plot, the projection of said face into the $H_I$--$f_a$ plane corresponds to the anthropic window displayed in \reffig{fig:parameter:space}. The shading is calibrated to the axion density; this is clear for the right plot but is also true for the left plot. We see that the parameter space does not depend sensitively on $\Omega_a$ in the range $\Omega_a \in [\Omega_c,0.1\,\Omega_c]$; for smaller densities $H_I$ starts to grow rapidly and eventually we reach the regime $\langle \theta_a \rangle \ll \sigma_{\theta_a} \ll 1$ where the isocurvature bound can no longer be fulfilled. This gives another bound in the anthropic window, although a rather uninteresting one since the axion density has become totally negligible at that point.}
\label{fig:anthropic:3d}
\end{center}
\end{figure}

If we allow the axion to contribute only a fraction of the dark matter content to the universe, the parameter space in the anthropic region becomes truly 3-dimensional. For $\theta_a<\theta_a(f_a)$ the axion density drops, and smaller effective angles are in one-to-one correspondence with $\Omega_a<\Omega_c$ for fixed $f_a$. Note that the relation $\Omega_a=\Omega_a(f_a,H_I,\langle \theta_a \rangle)$ can be inverted and allows us to trade $\langle \theta_a \rangle$ for $\Omega_a$. In \reffig{fig:anthropic:3d} we plot the available parameter space in the anthropic window.

It was somewhat surprising that $f_a(\theta_a)$ varies very slowly as $\pi-\theta_a \to 0$; indeed in that regime the axion field behaves like an inflaton, and one might have expected that $f_a \to 0$ rather fast. This is in agreement with the earlier numerical work by \cite{turner:axion:cosmology} and analytic computations \cite{lyth:axion:inflation:fluctuations:1,strobl:weiler:axion:anharmonic,visinelli:gondolo:axion}. It follows from this that the axion potential is not flat enough to support a prolonged inflationary period. In this regime, the fine-tuning in $\theta_a$ to many digits is reminiscent of the fine-tuning for the couplings of the inflationary potential to produce sufficient e-folds.

\subsection{Axion strings}

For completeness, let us take a closer look at the ``classic axion window'', that is, the thermal scenario in which the universe reheats to temperatures above the PQ symmetry breaking scale, $T > f_a$.  There are a number of misconceptions in the literature about the importance of misalignment production of axions from this thermal state.  At high temperature, the thermal axion distribution $\theta_a$ will fluctuate randomly in time and space around the circle $0 \rightarrow 2\pi$. An rms average with $\langle \theta \rangle= \pi/\sqrt{3}$ away from the minimum at $\theta_a=0$ is assumed to be the thermal initial condition for a misalignment zero momentum mode (\ref{eq:misalignment:bound}) which begins to oscillate when the axion mass switches on. Unfortunately, this generically underestimates axion production in the thermal scenario because it does not quantitatively account for the full spectrum of axion momentum states which are excited by the time of the mass `switch on'.

Axion production in the thermal scenario is in fact dominated by radiation from axion strings which inevitably form via the Kibble mechanism \cite{kibble:mechanism} when the $U_{\mathrm{PQ}}(1)$ symmetry breaks. These strings radiate axions during their scale invariant evolution on all subhorizon scales and then annihilate and disappear at axion mass switch on (as we will explain in more detail below). The important point is that axion strings at late times can effectively radiate low frequency axions in the range $H < \omega < m_a$ (for $f_a \ll 10^{16} \units{GeV}$). When the mass switches on, it is fairly straightforward to show that the resulting density of these non-relativistic axions inevitably exceeds those naively estimated from the zero momentum mode. This discussion follows the review given in ref.~\cite{battye:shellard:axion:string:1994b,battye:shellard:axion:string:1994b:error,Battye:1997jk}, updating the string constraint using new data. Note that there has been a long and intermittent history to the study of axion (or global) strings with numerical simulations performed by several groups \cite{battye:shellard:axion:string:1994b,hagmann:chang:sikivie:axion:string,moore:shellard:martins:abelian:higgs:string:network,yamaguchi:yokoyama:global:strings:lagrangian}. While there is general agreement on what happens on microphysical scales, there has been a variety of different interpretations offered when extrapolated over many orders of magnitude to cosmological scales. Here, we affirm the basic picture outlined originally in ref.~\cite{Vilenkin:1982ks}, while an alternative view can be found in ref.~\cite{Sikivie:2006ni}, an interpretation for which an earlier critique was offered in ref.~\cite{Battye:1997jk}.

Axion strings are global strings which possess strong long-range forces, with a (renormalised) energy per unit length given by
\begin{equation}
\mu \approx 2\pi f_a^2 \ln (L/\delta)\,,
\end{equation}
where $\delta \sim f_a^{-1}$ is the string core width and the typical radius of curvature is $L\sim t$.   At late times near axion mass switch on, the logarithm $\ln (L/\delta)\approx 70$ for $f_a \sim 10^{11}$GeV. Here, we essentially have a renormalisation of the bare string energy density $\mu_0 = 2\pi f_a^2$ by radially integrating out the effect of the axion field that winds by $2\pi$ around the string. Despite the nomenclature, with such a large $\mu \gg \mu_0$, the axion string is in fact highly localised with more than $95\%$ of its energy in only $0.1\%$ of the volume through which it traverses; it behaves to lowest order like a (local) Nambu string.

The evolution of a network of cosmic strings is non-pathological (whether local or global), because their evolution is scale-invariant  with their curvature radius growing as $L = \gamma t$ and the string density scaling as $\rho = \zeta \mu/t^{2}$, that is, in proportion to the background cosmological density. In the radiation era, we have $\zeta \equiv \gamma^{-2} \approx 13$, as determined by large-scale numerical simulations \cite{bennett:bouchet:axion:string:1,allen:shellard:axion:string}. The string network consists of both a population of long strings, that carry most of the energy, and small (subhorizon) loops which are created through long string reconnections,
\begin{equation}
 \rho_\mathrm{str} = \rho_\infty + \rho_\ell \equiv \frac{\mu \zeta}{t^2} + \mu\int \ell \,n(\ell, t)d\ell\,,
\end{equation}
where $n(\ell, t)d\ell$ is the number density of loops in the range $\ell$ to $\ell +d\ell$ at the time $t$.
By conservation of energy for the string network, one can calculate the loop number density (see the discussion in e.g.\~ref~\cite{vilenkin:shellard:defects}):
\begin{equation}
n(\ell, t ) d\ell =\frac{ \zeta g (1-\langle v^2\rangle)\alpha ^{1/2} }{\ell^{5/2}t^{3/2}}\,,\label{eq:loopdist}
\end{equation}
where $\alpha$ describes the typical loop size at creation $\ell = \alpha t$, the rms string velocity is $\langle v^2\rangle^{1/2}\approx 0.65$, and the relativistic boost factor $g\approx 1/\sqrt{2}$ accounts for centre-of-mass kinetic energies at loop creation (with $\nu_0\equiv g\langle v^2\rangle\approx 0.28$ reliably measured in simulations). Significant uncertainty remains concerning the loop creation size $\alpha$ given the complex nature of small scale structure on strings and the limited dynamic range available in simulations. The most recent and largest scale simulations indicate that the largest loops are produced on length scales  $\alpha \approx 0.1$ and below, but with a fairly flat loop production function which extends down to smaller scales \cite{Olum:2006ix}.
 
Loops oscillate periodically and decay fairly rapidly into axions. Axion radiation is primarily emitted in low frequencies, that is, in the lowest harmonics of the fundamental frequency of the loop oscillation $\omega_0 = 4\pi/L$ (the loop period is $T=L/2$). We note that axion radiation by strings can be treated very similarly to radiation into gravitational waves \cite{vilenkin:shellard:defects}. Decomposing the loop radiation power $P$ per oscillation into a spectrum $P_n$ for each harmonic $n\omega_0$ we expect
\begin{equation}
P = \sum_n P_n\,,\qquad P_n\propto n^{-q}~~(n\gg 1), \quad \hbox{with} ~~ q\ge 4/3\,.
\end{equation}
Here, the spectral index $q=4/3$ can be calculated analytically for loops with cusps, but realistically this is expected to be greater than $4/3$ because of radiative backreaction, with an effective maximum $n_*$. Given the dominance of the low harmonics in these spectra, the following results are relatively insensitive to the details at large $n$. The typical integrated power per oscillation is described by $\Gamma_a$, defined in the continuum limit with a radiation spectrum $g(\ell \omega)$ as
\begin{equation}
P = \Gamma_a f_a^2 = f_a^2 \int g(x) dx  \quad \hbox{with} ~~\frac{dP_\ell (\omega)}{d\omega} = f_a^2 \ell g(\ell\omega)\,. \label{eq:loopspectrum}
\end{equation}
The typical loop radiation rate $\Gamma_a$ can be estimated analytically for specific loop trajectories (see, for
example, ref.~\cite{Allen:1994bs}), but it is measured numerically from simulation loop trajectories to be $\Gamma_a \approx 65$ \cite{Allen:1991bk}.

\begin{figure}[tbp]
\begin{center}
\includegraphics[width=\figwidth,clip=true,trim=0mm 0mm 15mm 10mm]{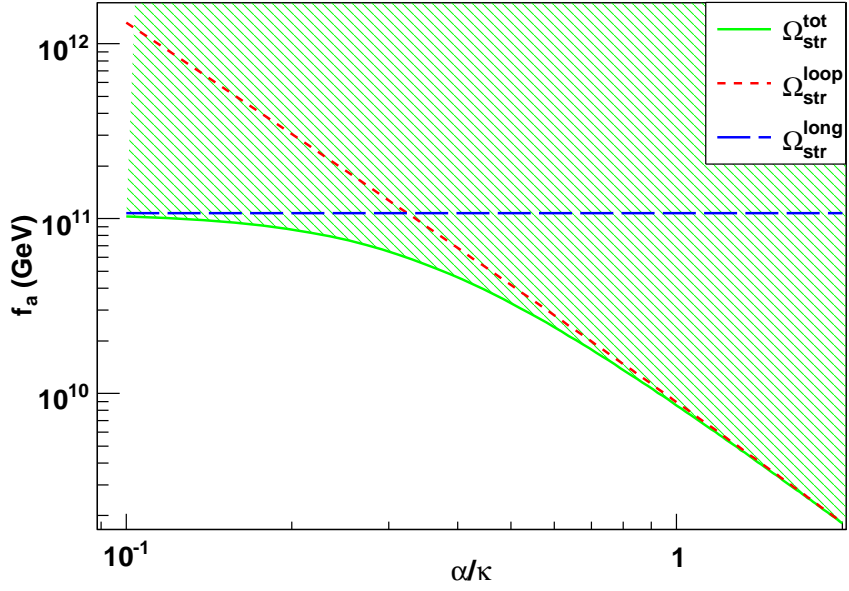}
\caption{Effect of the loop creation ratio $r\equiv\alpha/\kappa$ on the dark matter axion constraint. If $\alpha(t)$ the loop creation size at time $t$ is larger than $\kappa(t)$, the loop radiation backreaction scale, then the constraint is stronger and conversely for $\alpha < \kappa$. Note that in the second case with $\alpha \ll \kappa$, the dominant contribution arises from direct long string radiation, which again exceeds misalignment production.}
\label{fig:string:uncertainties}
\end{center}
\end{figure}

Due to these radiative losses into axions, a loop will shrink linearly from its original size at creation $\ell_{\mathrm i} = \alpha t_{\mathrm i}$ as
\begin{equation}
\ell = \ell_{\mathrm i} - \kappa (t - t_{\mathrm i})\,\quad \hbox{with} ~~\kappa  = \Gamma_a / \mu \approx \Gamma_ a /2\pi f_a^2\ln(\ell/\delta)\,.
\end{equation}
Given $\ln(\ell/\delta)\approx 70$ for the energy scales and times of interest, this means that the loop backreaction rate $\kappa \approx 0.15$ and a loop will oscillate about 10-15 times before vanishing. This loop decay will modify and cut-off the loop distribution given above (\ref{eq:loopdist}), becoming 
\begin{equation}
n(\ell, t ) d\ell =\frac{ \zeta \nu_0\alpha ^{1/2} (1+\kappa/\alpha)^{3/2}}{(\ell+\kappa t)^{5/2}t^{3/2}}\,.\label{eq:loopraddist}
\end{equation}
Given  the spectral assumptions (\ref{eq:loopspectrum}), we can integrate over the loop distribution \ref{eq:loopraddist} to obtain a spectrum for the number density of axions $n_a$ (for $\alpha\lesssim \kappa$) \cite{Battye:1997jk}
\begin{equation}
\frac{dn_a}{d\omega} =\frac {1}{\omega} \frac{d\rho_a}{d\omega}  = \frac{ 4\Gamma_a\zeta \nu_0\alpha ^{1/2} (1+\kappa/\alpha)^{3/2}}{3\omega^2\kappa^{3/2}t^2} \left [ 1 - \left(1+\frac{\alpha}{\kappa}\right)^{-3/2}\right]\,.
\end{equation}
Integrating down to the lowest frequencies $\omega = 4\pi/\alpha t$ emitted at a time $t$, we obtain the total axion number density \cite{Battye:1997jk}
\begin{equation}
n_a=\frac{ \Gamma_a\zeta \nu_0 }{3\pi t} \left [  \left(1+\frac{\alpha}{\kappa}\right)^{3/2}-1\right]\,,\label{eq:loopaxiondensity}
\end{equation}
where the prefactor has a numerical value $\Gamma\zeta\nu_0/3\pi \approx 31$ with moderate uncertainties  ($\pm30\%$). The most important uncertainty in the expression (\ref{eq:loopaxiondensity}) is clearly the loop size parameter $\alpha$ measured relative to the backreaction rate $\kappa$. In \reffig{fig:string:uncertainties}, we see the strong dependence of the axion string bound on the ratio $\alpha/\kappa$ having imposed the dark matter constraint $\Omega_a \le 0.23$. Recent string simulations suggest $\alpha \lesssim 0.1$ implying $\alpha/\kappa \lesssim 0.7$, but what is the lower limit or, rather, the appropriate range for $\alpha$?

The lower cutoff for the loop production scale $\alpha$ depends on the nature of small-scale structure on long strings, since it is unlikely that loops can be produced on scales below which the string network becomes smooth. Numerical and analytic studies of long string backreaction indicate that it is somewhat weaker than loop backreaction because of geometric effects and the typical amplitude and velocity of long string oscillations. The loop radiation rate $\kappa$ is then replaced by the long string rate \cite{Battye:1995hw} 
\begin{equation}
\kappa_\infty \sim \frac{\pi^2}{8}\ln(t/\delta) ~\sim~ 0.02\,,\label{eq:loopcutoff}
\end{equation}
(refer also to related studies of gravitational waves from long strings in ref.~\cite{Olum:1999sg}). The expression (\ref{eq:loopcutoff}) indicates a lower limit on the loop production size $\ell \gtrsim \kappa_\infty t$, thus with the loop size ratio lying in a fairly narrow range $0.1 \lesssim \alpha/\kappa \lesssim 0.7$. Nevertheless, it is the larger loops with $\alpha/\kappa \sim 0.5$ which are expected to dominate the axion contribution, given the nature of the spectral weighting underlying (\ref{eq:loopaxiondensity}).

There is also a significant contribution coming from direct radiation from long strings which, given (\ref{eq:loopcutoff}), has a spectral radiation rate \cite{Battye:1997jk,Battye:1995hw} 
\begin{equation}
\frac{dn_a^\infty}{d\omega} =\frac {1}{\omega} \frac{d\rho_a}{d\omega}  \approx \frac{\pi^3 f_a^2\zeta}{8\kappa_\infty\omega^2 t^2}\,.\label{eq:loopradpower}
\end{equation}
This integrates to produce the additional axion number density
\begin{equation}
n_a^\infty=\frac{ \pi^2 f_a^2 \zeta}{32 t} \,, \label{eq:longaxiondensity}
\end{equation}
where we have assumed that the radiation spectrum is peaked at the backreaction scale $\omega \approx 4\pi/\kappa_\infty t$ (second harmonic). This assumption is certainly a conservative underestimate because there will be significant radiation from the strings on the correlation length and above, up to the horizon scale $\omega \gtrsim H$, i.e.\ effectively contributing more non-relativistic axions at mass `switch on'.

Taken together, the sum of the axions produced by loops and long strings is shown in \reffig{fig:string:uncertainties}, indicating that this is a significant stronger constraint than the misalignment estimate even if loops were very small $\alpha/\kappa \ll 0$. Given that present numerical simulations indicate that a significant proportion of loops are produced on the largest scales in the available range $0.1 \lesssim \alpha/\kappa \lesssim 0.7$, we take the effective value $\alpha/\kappa = 0.5\pm0.2$ to obtain an updated axion string constraint
\begin{equation}
 f_a~\lesssim~3.2^{+4}_{-2}\times10^{10} \units{GeV} \, ,\qquad m_a ~\gtrsim~ 0.20 ^{+0.2}_{-0.1} \units{meV}\,,\label{eq:axionstringconstraint}
\end{equation}
that is, we predict that if thermal axions are the predominant dark matter then they will have a mass near
$m_a \approx 200\,\mu$eV (revising slightly upward previous estimates from strings \cite{Battye:1997jk}). We emphasise that the axion string constraint (\ref{eq:axionstringconstraint}) is a conservative bound which is an order of magnitude stronger than the misalignment estimate (\ref{eq:misalignment:bound}).

The axion string contribution is computed around the period of mass `switch on', defined by $m_a \approx 3H$. This induces a dependence on the axion mass. It turns out that the differences are fairly small, i.e.\ on the $1\%$ level. Given the large uncertainties in the axion string computation, the axion mass dependence is certainly negligible.

Finally, we comment on additional contributions, uncertainties and alternative views of the thermal scenario. At the QCD phase transition, the tilting axion potential causes the axion field $\theta$ to relax toward the minimum $\theta=0$, assisted on long wavelengths ($\omega\sim H$) by Hubble damping -- as for the zero momentum mode in inflationary scenarios. However, the presence of a string implies a topological obstruction to this relaxation, so the field instead localises its variations $0\rightarrow 2\pi$ into domain walls connecting strings to others of the opposite orientation \cite{Sikivie:1982qv, Vilenkin:1982ks}. This correlation of strings within the network, as well as the intercommuting properties of strings colliding with walls, leads to the rapid demise of the hybrid network as demonstrated numerically \cite{Shellard:1986in, Shellard:1987}. Self-intersections of loops bounding domain walls are inevitable until the system breaks up into tiny loops $\ell \lesssim m_a^{-1}$ and wall tension effects become negligible. These loops can decay by radiating (massive) axions and gravitational waves. Estimates of the axion number density $n_a^{\mathrm{dw}}$ produced by this process are crude but suggest a contribution comparable only to that from misalignment \cite{Chang:1998tb}. Axion production through hybrid network formation deserves more detailed analysis particularly for nonrelativistic modes with $H \lesssim \omega \lesssim m_a$.

The key uncertainty in the string constraint above remains the typical loop production size $\alpha$ for a global string network. This is difficult to probe numerically for two reasons. First, Nambu string simulations are progressing in resolution but they do not at present include the realistic radiative backreaction necessary for describing global strings. The understanding of $\alpha$ is improving but how it cuts off on small scales must be estimated analytically. Secondly, alternative field theory simulations do not have the dynamic range needed to
address the loop production issue; with $\ln(t/\delta) \sim 3 \ll 70$, simulated global strings remain in a strongly damped regime unlike cosmological axion strings. It has been suggested that the spectrum of loop and long string radiation could be `flat' (see, for example, \cite{Sikivie:2006ni} and references therein), rather than dominated by the lowest harmonics as in the standard picture presented above (\ref{eq:loopspectrum}). The proposed `flat' spectrum entails producing equal radiation power over many orders of magnitude $t \lesssim \omega \lesssim f_a$ (up to 30), effectively suppressing the importance of the long wavelength modes on which the bound (\ref{eq:axionstringconstraint}) is based. The `flat' spectrum yields an axion string constraint roughly matching the misalignment bound (\ref{eq:misalignment:bound}) \cite{hagmann:chang:sikivie:axion:string}, but a detailed critique of this scenario is offered elsewhere \cite{Battye:1997jk}.

\section{Conclusion}

We have presented a temperature-dependent axion mass, based on instanton methods, that is valid for all temperatures. The transition between the high- and low-temperature regime is well-motivated and computed within the same model, the IILM, in contrast to some ad hoc procedure to connect the two.

Although the IILM does not explain confinement, chiral symmetry restoration is incorporated and the model can be expected to give qualitatively correct results for the axion mass; this relies on the fact that it is related to the QCD topological susceptibility which in turn is a chiral quantity. We note that chiral symmetry restoration is indeed seen in the IILM, although at a slightly lower temperature. Given the discovery of the more general KvBLL calorons, that may play an important role in the confinement/deconfinement transition, we expect to improve our understanding of the axion potential in the future by incorporating these new degrees of freedom into the IILM.

Using this new axion mass, we solved numerically the axion evolution equations in the concordance FRW cosmology. It turns out that the analytic approximations used previously differ by a factor of $2-3$. This is unexpectedly good agreement, considering the crude determination of the axion mass within the dilute gas approximation. We believe it to be the result of a coincidence, that the extrapolation of the high temperature DGA axion mass fairly closely follows the full IILM result around the phase transition. Conversely, this correspondence can be interpreted as evidence that the axion mass determination is fairly robust. This is also seen from the rather small differences between the numerical results between the IILM mass and the lattice inspired mass. 

We want to draw to attention that the IILM is a model of QCD, and as such it needs to be checked against lattice data, say. In that light, there remains the possibility that the true high temperature axion mass is different from the IILM prediction. This could lead to qualitatively different results: On the one hand, a softer decay, as seen in gluodynamics, will lead to weaker constraints. A more abrupt mass switch on, on the other hand, would tighten the constraints and potentially close the classic axion window. The ideas of molecule formation within the IILM, and the subsequent stronger suppression of the axion mass in the plasma phase, have initially prompted this investigation. Within the current IILM this is not realised, but it is not ruled out either. The lattice community is performing realistic QCD simulations directly at the physical quark masses, and we can expect a reliable axion mass determination to follow from that data in future.

To get an accurate estimate across the whole $f_a$ axis, we included the correct temperature dependence between the scale factor and the plasma temperature which follows from the conservation of entropy. To that end we computed the full phase-space integral to get the temperature dependence for the effective degrees of freedom $g_{*.S}$ and $g_{*,R}$, following closely \cite{coleman:roos:geff}. Additionally, we provide accurate fitting formulas, which to our knowledge have not been presented in the literature previously.

In the classic axion window, where the PQ symmetry breaks only after inflation, a quantitative analysis of the misalignment bound yields $f_a < 2.8(\pm2) \times 10^{11}\units{GeV}$, $m_a>21(\pm 2) \units{\mu eV}$. While the misalignment axion is a useful benchmark for comparing alternative estimates, it is far exceeded in density by axions from string radiation which yields the more stringent string bound $f_a~\lesssim~3.2^{+4}_{-2} \times 10^{10} \units{GeV}$. Experimental searches for a thermal dark matter axion should focus on masses around $m_a \sim 200 \units{\mu eV}$.

The anthropic axion window, which entangles $\theta_a$ and $f_a$, is constrained by investigating the production of isocurvature perturbations. The rather small contribution of these to the CMBR power-spectrum places strong bounds on $f_a$ and the inflationary scale $H_I$. This bound is strong enough so that quantum fluctuations are totally negligible, i.e.\ $\sigma_a \ll \langle \theta_a \rangle$, for almost all axion angles. The anthropic window allows a dark matter axion that can have any initial misalignment angle, provided the inflationary scale is low enough. In light of naturalness in fundamental theory, most interest has focused on the constraints for large $f_a$.

Nevertheless, taking the whole of the anthropic range seriously reveals some interesting possibilities. In particular, we investigated the region with large misalignment angles, especially $\theta_a \to \pi$. In this regime the axion potential can no longer be described by a parabola, and we need to take the anharmonic effects into account. A self-consistent solution follows for the regime with small fluctuations, in which case we can Taylor-expand the axion potential around $\langle \theta_a \rangle$; the anharmonic effects are then encoded in the anharmonic dependence on $\langle \theta_a \rangle$. These effects lead to smaller isocurvature perturbations as compared to the harmonic case and, consequently, lead to weaker constraints on the inflationary scale.

Fine-tuning the initial misalignment angle to $\pi$, we find that the quantum fluctuations must be even more stringently constrained. If the axion provides the dominant component to the dark matter content of the universe and if we take at face value an inflationary period after the spontaneous breaking of the PQ symmetry, then this places the strongest constraint on $f_a$ and $H_I$ at large initial misalignment angle. We note the intriguing possibility of a dominant dark matter axion with a mass $m_a \sim 200 \units{\mu eV}$ which is consistent with either the thermal or the inflationary scenarios. This model independence provides extra motivation for experimental searches around this mass range. Anthropic tuning near $\theta\approx \pi$ allows inflationary dark matter axions ($\Omega_a = 0.23$) to have masses as high as $m_a\le 1 \units{meV}$, but no higher because quantum fluctuations restrict the fine tuning (for $H_I \ge 10^4 \units{GeV}$).

We note that the isocurvature and quantum fluctuation constraints become weaker if the axion is not the dominant dark matter candidate, and we displayed the ensuing 3-dimensional parameter space. 

Finally, the dependence of the cosmological constraints on the axion mass is only a secondary effect for the isocurvature constraint and the axion string contribution. In the former case the examined parameter space is so large that only significant modifications play a role, such as the change from a harmonic to an anharmonic potental; in the latter case, the uncertainties inherent in the computation easily exceed those from the axion mass dependence.

\section*{Acknowledgements}

We are very grateful for many informative discussions with P. Faccioli on the instanton liquid and R. Battye on axion strings. Simulations were performed on the COSMOS supercomputer (an Altix 4700) which is funded by STFC, HEFCE and SGI. OW was supported by STFC grant PPA/S/S2004/03793 and an Isaac Newton Trust European Research Studentship. E.P.S. S. was supported by STFC grant ST/F002998/1 and the Centre for Theoretical Cosmology.

\appendix
\section{Effective degrees of freedom}
\label{app:geff:fit}

The entropy and radiation density is well approximated by counting only the relativistic degrees of freedom. When the temperature of the plasma drops below the mass, the particle's contribution to the entropy and radiation density drops to zero fairly rapidly. To compute these threshold effects correctly, the relevant phase-space integrals have to be evaluated, which is not possible in closed form in general. The effective (relativistic) degrees of freedom are defined by factoring out the contribution from a single massless degree of freedom, of temperature $T$, from the phase space integrals; this leads to the expression (\ref{eq:geff:S}) and (\ref{eq:geff:R}). Since the photon field will be the only relativistic field at sufficiently low temperatures, it is used to set the temperature of the plasma, i.e.\ $T=T_\gamma$.

If a species decouples, it will no longer be in contact with the heat bath, and its temperature will simply redshift due to the Hubble expansion. This is in contrast to the evolution of the plasma temperature which follows from entropy conservation. Thus, if a species becomes non-relativistic, it transfers its entropy only to those particles that are still in equilibrium with the plasma, and reheats the plasma since the number of effective degrees of freedom has decreased. This mechanism leads to the different temperatures $T_i$ for the decoupled fields.

In standard cosmology this happens to the neutrinos which decouple before $e^\pm$-annihilation, and so have a slightly lower temperature than the photons, $T_\nu=\left(\frac{4}{11}\right)^{1/3} T_\gamma$; this result is valid at $T\ll1\units{MeV}$. For a general temperature $T$, the neutrino temperature is given by
\begin{equation}
 T_\nu=T \left(\frac{g_{*,S}(T)}{g_{*,S}(T^d_\nu)}\right)^{1/3} ,\; \mbox{if } T<T^d_\nu\,, \label{eq:neutrino:temperature}
\end{equation}
where $T^d_\nu \approx 1\units{MeV}$ is the neutrino decoupling temperature. It is a consequence of the fact that for decoupled (relativistic) particles such as neutrinos $T^i_\nu/T^f_\nu = a_f/a_i$, where the ratio of the scale factors follows from the adiabatic evolution of the universe. In fact, (\ref{eq:neutrino:temperature}) is an implicit equation in $T_\nu$ since $g_{*,S}(T)$ depends on it. With only neutrinos decoupling, it is easy to solve, and the result is (\ref{eq:neutrino:temperature}) with the substitution $g_{*,S} \to g_{*,S} - g_\nu$. In order to reproduce the correct analytic ratio for low temperature, we have set $T^d_\nu=5\units{MeV}$. This then leads to \reffig{fig:geff}.

Evaluation of the exact numerical integration is fast but too slow for use in the system of ODE's that needs to be solved for the axion dynamics. We have approximated the exact result by fits that are sums of smoothed out step functions in log-log space. These fits are plotted in \reffig{fig:geff} and are generically accurate below the $1\%$ level, except at the QCD phase transition and the $e^\pm$-annihilation where the accuracy is only around $4\%$. The fits are given by
\begin{equation}
 g_{*,i}=\exp\left[a^i_0+\sum_{j=1}^5 a^i_{j,1}\left(1.0+\tanh\frac{t-a^i_{j,2}}{a^i_{j,3}}\right)\right],\;t=\log\frac{T}{1\units{GeV}}\,.
\end{equation}
The corresponding parameters are
\begin{equation}
 \begin{array}{c|c|c|c|c|c}
  j & 1 & 2 & 3 & 4 & 5 \\\hline\hline
  a^R_0 & \multicolumn{5}{c}{1.21}\\\hline
  a^R_{j,1} & 0.572 & 0.330 & 0.579 & 0.138 & 0.108 \\\hline
  a^R_{j,2} & -8.77 & -2.95 & -1.80 & -0.162 & 3.76 \\\hline
  a^R_{j,3} & 0.682 & 1.01 & 0.165 & 0.934 & 0.869 \\\hline\hline
  a^S_0 & \multicolumn{5}{c}{1.36}\\\hline
  a^S_{j,1} & 0.498 & 0.327 & 0.579 & 0.140 & 0.109 \\\hline
  a^S_{j,2} & -8.74 & -2.89 & -1.79 & -0.102 & 3.82 \\\hline
  a^S_{j,3} & 0.693 & 1.01 & 0.155 & 0.963 & 0.907
 \end{array}
\end{equation}



\begin{thebibliography}{10}

\bibitem{Allen:1994bs}
B.~Allen, P.~Casper, and A.~Ottewill.
\newblock {Analytic results for the gravitational radiation from a class of
  cosmic string loops}.
\newblock {\em Phys. Rev.}, D50:3703--3712, 1994.

\bibitem{Allen:1991bk}
B.~Allen and E.~P.~S. Shellard.
\newblock {Gravitational radiation from cosmic strings}.
\newblock {\em Phys. Rev.}, D45:1898--1912, 1992.

\bibitem{allen:shellard:axion:string}
B.~Allen and E.P.S. Shellard.
\newblock {COSMIC STRING EVOLUTION: A NUMERICAL SIMULATION}.
\newblock {\em Phys. Rev. Lett.}, 64:119--122, 1990.

\bibitem{amsler:2008:pdg}
C.~Amsler et~al.
\newblock {Review of particle physics}.
\newblock {\em Phys. Lett.}, B667:1, 2008.

\bibitem{aoki:fodor:katz:szabo:transition:temperature}
Y.~Aoki, Z.~Fodor, S.D. Katz, and K.K. Szabo.
\newblock {The QCD transition temperature: results with physical masses in the
  continuum limit}.
\newblock {\em Phys.Lett. B}, 643:46--54, 2006.

\bibitem{bae:huh:kim:axion}
K.~Bae, J.~Huh, and J.~Kim.
\newblock {Update of axion CDM energy}.
\newblock {\em JCAP}, 0809:005, 2008.

\bibitem{baker:electric:dipole:neutron}
C.~A. Baker et~al.
\newblock {An improved experimental limit on the electric dipole moment of the
  neutron}.
\newblock {\em Phys. Rev. Lett.}, 97:131801, 2006.

\bibitem{baluni:cp:violation:qcd}
V.~Baluni.
\newblock {CP Violating Effects in QCD}.
\newblock {\em Phys. Rev.}, D19:2227--2230, 1979.

\bibitem{battye:shellard:axion:string:1994b}
R.~Battye and E.P.S. Shellard.
\newblock {Axion String Constraints}.
\newblock {\em Phys. Rev. Lett.}, 73:2954--2957, 1994.

\bibitem{battye:shellard:axion:string:1994b:error}
R.~Battye and E.P.S. Shellard.
\newblock {Axion String Constraints}.
\newblock {\em Phys. Rev. Lett.}, 76:2203, 1996.

\bibitem{Battye:1995hw}
R.~A. Battye and E.~P.~S. Shellard.
\newblock {Radiative back reaction on global strings}.
\newblock {\em Phys. Rev.}, D53:1811--1826, 1996.

\bibitem{Battye:1997jk}
R.~A. Battye and E.~P.~S. Shellard.
\newblock {Recent perspectives on axion cosmology}.
\newblock 1997.

\bibitem{beltran:garcia-bellido:lesgourges:isocurvature}
M.~Beltran, J.~Garcia-Bellido, and J.~Lesgourgues.
\newblock {Isocurvature bounds on axions revisited}.
\newblock {\em Phys. Rev.}, D75:103507, 2007.

\bibitem{bennett:bouchet:axion:string:1}
D.~Bennett and F.~Bouchet.
\newblock {High resolution simulations of cosmic string evolution. 1. Network
  evolution}.
\newblock {\em Phys. Rev. D}, 41:2408, 1990.

\bibitem{burns:axion:isocurvature}
S.~Burns.
\newblock {Isentropic and isocurvature axion perturbations in inflationary
  cosmology}.
\newblock 1997.

\bibitem{callan:dashen:gross:theory:strong:interactions}
C.G. Callan, R.F. Dashen, and D.J. Gross.
\newblock {Toward a theory of the strong interactions}.
\newblock {\em Phys.Rev. D}, 17:2717--2763, 1978.

\bibitem{callan:dashen:gross:theory:hadronic}
C.G. Callan, R.F. Dashen, and D.J. Gross.
\newblock {A theory of hadronic structure}.
\newblock {\em Phys.Rev. D}, 19:1826--1855, 1979.

\bibitem{Chang:1998tb}
Sanghyeon Chang, C.~Hagmann, and P.~Sikivie.
\newblock {Studies of the motion and decay of axion walls bounded by strings}.
\newblock {\em Phys. Rev.}, D59:023505, 1999.

\bibitem{cheng:strongCP:report}
H.-Y. Cheng.
\newblock {The Strong CP Problem Revisited}.
\newblock {\em Phys. Rept.}, 158:1, 1988.

\bibitem{coleman:roos:geff}
T.~Coleman and M.~Roos.
\newblock {Effective degrees of freedom during the radiation era}.
\newblock {\em Phys. Rev.}, D68:027702, 2003.

\bibitem{crewther:divecchia:veneziano:witten:electric:dipole:moment}
R.~Crewther, P.~Di Vecchia, G.~Veneziano, and E.~Witten.
\newblock {Chiral Estimate of the Electric Dipole Moment of the Neutron in
  Quantum Chromodynamics}.
\newblock {\em Phys. Lett.}, B88:123, 1979.

\bibitem{dabholkar:quashnock:axion:string}
A.~Dabholkar and J.~Quashnock.
\newblock {PINNING DOWN THE AXION}.
\newblock {\em Nucl. Phys.}, B333:815, 1990.

\bibitem{davis:axion:string:1985}
R.~Davis.
\newblock {Goldstone bosons in string models of galaxy formation}.
\newblock {\em Phys. Rev. D}, 32:3172, 1985.

\bibitem{davis:axion:string:1986}
R.~Davis.
\newblock {Cosmic axions from cosmic strings}.
\newblock {\em Phys. Lett. B}, 180:225, 1986.

\bibitem{davis:shellard:axion:string}
R.~Davis and E.P.S. Shellard.
\newblock {Do axions need inflation}.
\newblock {\em Nucl. Phys. B}, 324:167, 1989.

\bibitem{dine:fischler:srednicki:axion}
M.~Dine, W.~Fischler, and M.~Srednicki.
\newblock {A simple solution to the strong CP problem with a harmless axion}.
\newblock {\em Phys. Lett. B}, 104:199, 1981.

\bibitem{dunne:hur:lee:min:instanton:determinant:mass}
G.V. Dunne, J.~Hur, C.~Lee, and H.~Min.
\newblock {Calculation of QCD Instanton Determinant with Arbitrary Mass}.
\newblock {\em Phys.Rev. D}, 71:085019, 2005.

\bibitem{diakonov:instanton:nonzero:T}
D.I. Dyakonov and A.D. Mirlin.
\newblock {Instanton vacuum at non-zero temperatures}.
\newblock {\em Phys. Lett. B}, 203:299--304, 1988.

\bibitem{diakonov:instanton:variational}
D.I. Dyakonov and V.Yu. Petrov.
\newblock {Instanton-based vacuum from the Feynman variational principle}.
\newblock {\em Nucl. Phys. B}, 245:259--292, 1984.

\bibitem{diakonov:instanton:quarks}
D.I. Dyakonov and V.Yu. Petrov.
\newblock {A theory of light quarks in the instanton vacuum}.
\newblock {\em Nucl. Phys. B}, 272:457--489, 1986.

\bibitem{fox:pierce:scott:axion:cosmology}
P.~Fox, A.~Pierce, and S.~Thomas.
\newblock {Probing a QCD string axion with precision cosmological
  measurements}.
\newblock 2004.

\bibitem{fugleberg:halperin:zhitnitsky:domain:walls:theta:qcd}
T.~Fugleberg, I.~Halperin, and A.~Zhitnitsky.
\newblock {Domain walls and theta dependence in {QCD} with an effective
  Lagrangian approach}.
\newblock {\em Phys. Rev.}, D59:074023, 1999.

\bibitem{gabadadze:shifman:vacuum:structure:qcd}
G.~Gabadadze and M.~Shifman.
\newblock {Vacuum structure and the axion walls in gluodynamics and QCD with
  light quarks}.
\newblock {\em Phys. Rev.}, D62:114003, 2000.

\bibitem{gabadadze:shifman:qcd:vacuum:axion}
G.~Gabadadze and M.~Shifman.
\newblock {QCD vacuum and axions: What's happening?}
\newblock {\em Int. J. Mod. Phys.}, A17:3689--3728, 2002.

\bibitem{gerhold:ilgenfritz:mueller_preussker:kvbll:gas:confinement}
P.~Gerhold, E.M. Ilgenfritz, and M.~M\"{u}ller-Preussker.
\newblock {An $SU(2)$ KvBLL caloron gas model and confinement}.
\newblock {\em Nucl. Phys. B}, 760:1--37, 2007.

\bibitem{gross:pisarski:yaffe:instantons:finite:T}
D.J. Gross, R.D. Pisarski, and L.G. Yaffe.
\newblock {QCD and instantons at finite temperature}.
\newblock {\em Rev. Mod. Phys.}, 53:43--80, 1981.

\bibitem{hagmann:chang:sikivie:axion:string}
C.~Hagmann, Sanghyeon Chang, and P.~Sikivie.
\newblock {Axion radiation from strings}.
\newblock {\em Phys. Rev.}, D63:125018, 2001.

\bibitem{hagmann:sikivie:axion:string}
C.~Hagmann and P.~Sikivie.
\newblock {Computer simulations of the motion and decay of global strings}.
\newblock {\em Nucl. Phys. B}, 363:247, 1991.

\bibitem{halperin:zhitnitsky:qcd:theta}
I.~Halperin and A.~Zhitnitsky.
\newblock {Anomalous effective Lagrangian and theta dependence in {QCD} at
  finite N(c)}.
\newblock {\em Phys. Rev. Lett.}, 81:4071--4074, 1998.

\bibitem{halperin:zhitnitsky:qcd:axion:potential}
I.~Halperin and A.~Zhitnitsky.
\newblock {Axion potential, topological defects and CP-odd bubbles in {QCD}}.
\newblock {\em Phys. Lett.}, B440:77--88, 1998.

\bibitem{halperin:zhitnitsky:yang:mills:theta}
I.~Halperin and A.~Zhitnitsky.
\newblock {Can Theta/N dependence for gluodynamics be compatible with 2pi
  periodicity in Theta?}
\newblock {\em Phys. Rev.}, D58:054016, 1998.

\bibitem{hamann:hannestad:raffelt:wong:isocurvature}
J.~Hamann, S.~Hannestad, G.~Raffelt, and Y.~Wong.
\newblock {Isocurvature forecast in the anthropic axion window}.
\newblock {\em JCAP}, 0906:022, 2009.

\bibitem{hannestad:mirizzi:raffelt:thermal:axion}
S.~Hannestad, A.~Mirizzi, and G.~Raffelt.
\newblock {New cosmological mass limit on thermal relic axions}.
\newblock {\em JCAP}, 0507:002, 2005.

\bibitem{harari:sikivie:axion:string}
D.~Harari and P.~Sikivie.
\newblock {On the evolution of global strings in the early universe}.
\newblock {\em Phys. Lett. B}, 195:361, 1987.

\bibitem{hertzberg:tegmark:wilczek:axion}
M.~Hertzberg, M.~Tegmark, and F.~Wilczek.
\newblock {Axion Cosmology and the Energy Scale of Inflation}.
\newblock {\em Phys. Rev.}, D78:083507, 2008.

\bibitem{karsch:recent:partII}
F.~Karsch.
\newblock {Recent lattice results on finite temperature and density QCD, part
  II}.
\newblock {\em PoS}, LAT2007:015, 2007.

\bibitem{kibble:mechanism}
T.~W.~B. Kibble.
\newblock {Topology of Cosmic Domains and Strings}.
\newblock {\em J. Phys.}, A9:1387--1398, 1976.

\bibitem{kim:axion}
J.~Kim.
\newblock {Weak-interaction singlet and strong CP invariance}.
\newblock {\em Phys. Rev. Lett.}, 43:103, 1979.

\bibitem{kim:report}
J.~Kim.
\newblock {Light pseudoscalars, particle physics and cosmology}.
\newblock {\em Phys. Rep.}, 150:1, 1987.

\bibitem{kim:axion:quintessential}
J.~Kim.
\newblock {QCD axion and quintessential axion}.
\newblock 2003.

\bibitem{kim:axion:cdm}
J.~Kim.
\newblock {Axion as a CDM component}.
\newblock 2007.

\bibitem{turner:kolb:cosmology}
E.W. Kolb and M.S. Turner.
\newblock {\em {The Early Universe}}.
\newblock Westview Press, 1990.

\bibitem{komatsu:wmap5:cosmological:interpretation}
E.~Komatsu et~al.
\newblock {Five-Year Wilkinson Microwave Anisotropy Probe (WMAP)
  Observations:Cosmological Interpretation}.
\newblock {\em Astrophys. J. Suppl.}, 180:330--376, 2009.

\bibitem{kuster:raffelt:beltran:axions}
M.~Kuster, G.~Raffelt, and B.~Beltr{\'a}n, editors.
\newblock {\em Axions}.
\newblock Springer, 2008.

\bibitem{leutwyler:smilga:spectrum:dirac}
H.~Leutwyler and A.~Smilga.
\newblock {Spectrum of Dirac operator and role of winding number in QCD}.
\newblock {\em Phys. Rev. D}, 46:5607--5632, 1992.

\bibitem{lyth:axion:inflation:fluctuations:1}
D.~H. Lyth.
\newblock {Axions and inflation: Sitting in the vacuum}.
\newblock {\em Phys. Rev.}, D45:3394--3404, 1992.

\bibitem{mason:trottier:horgan:davies:lepage:quark:masses}
Q.~Mason, H.~Trottier, R.~Horgan, C.~Davies, and G.~Lepage.
\newblock {High-precision determination of the light-quark masses from
  realistic lattice QCD}.
\newblock {\em Phys. Rev.}, D73:114501, 2006.

\bibitem{masso:rota:zsembinszki:axion:thermal}
E.~Masso, F.~Rota, and G.~Zsembinszki.
\newblock {On axion thermalization in the early universe}.
\newblock {\em Phys. Rev.}, D66:023004, 2002.

\bibitem{mclerran:mottola:axion:sphaleron}
L.~McLerran, E.~Mottola, and M.~Shaposhnikov.
\newblock {SPHALERONS AND AXION DYNAMICS IN HIGH TEMPERATURE QCD}.
\newblock {\em Phys. Rev.}, D43:2027--2035, 1991.

\bibitem{moore:shellard:martins:abelian:higgs:string:network}
J.~N. Moore, E.P.S. Shellard, and C.J.A.P. Martins.
\newblock {On the evolution of Abelian-Higgs string networks}.
\newblock {\em Phys. Rev. D}, 65:023503, 2002.

\bibitem{nilles:quintaxion}
H.~P. Nilles.
\newblock {Hidden sector axions: Physics and cosmology}.
\newblock 2003.

\bibitem{Olum:1999sg}
Ken~D. Olum and J.~J. Blanco-Pillado.
\newblock {Radiation from cosmic string standing waves}.
\newblock {\em Phys. Rev. Lett.}, 84:4288--4291, 2000.

\bibitem{Olum:2006ix}
Ken~D. Olum and Vitaly Vanchurin.
\newblock {Cosmic string loops in the expanding universe}.
\newblock {\em Phys. Rev.}, D75:063521, 2007.

\bibitem{peccei:quinn:cp2}
R.~Peccei and H.~Quinn.
\newblock {Constraints imposed by CP conservation in the presence of
  pseudoparticles}.
\newblock {\em Phys. Rev. D}, 16:1791, 1977.

\bibitem{peccei:quinn:cp1}
R.~Peccei and H.~Quinn.
\newblock {CP conservation in the presence of pseudoparticles}.
\newblock {\em Phys. Rev. Lett.}, 38:1440, 1977.

\bibitem{pokorski:gauge}
S.~Pokorski.
\newblock {\em {Gauge Field Theories}}.
\newblock {Cambridge University Press}, 2000.

\bibitem{schafer:shuryak:instantons:qcd:review}
T.~Schafer and E.~Shuryak.
\newblock {Instantons in QCD}.
\newblock {\em Rev. Mod. Phys.}, 70:323--426, 1998.

\bibitem{Shellard:1986in}
E.~P.~S. Shellard.
\newblock {AXIONIC DOMAIN WALLS AND COSMOLOGY}.
\newblock In *Liege 1986, Proceedings, Origin and early history of the
  universe* 173-180. (see Conference Index).

\bibitem{Shellard:1987}
E.~P.~S. Shellard.
\newblock {\em Quantum Effects in the Early Universe}.
\newblock PhD thesis, University of Cambridge, 1987.

\bibitem{shifman:vainshtein:zakharov:cp}
M.~Shifman, A.~Vainshtein, and V.~Zakharov.
\newblock {Can confinement ensure natural CP invariance of strong
  interactions?}
\newblock {\em Nucl. Phys. B}, 166:493, 1980.

\bibitem{Sikivie:1982qv}
P.~Sikivie.
\newblock {Of Axions, Domain Walls and the Early Universe}.
\newblock {\em Phys. Rev. Lett.}, 48:1156--1159, 1982.

\bibitem{sikivie:yang:axion}
P.~Sikivie and Q.~Yang.
\newblock {Bose-Einstein Condensation of Dark Matter Axions}.
\newblock {\em Phys. Rev. Lett.}, 103:111301, 2009.

\bibitem{Sikivie:2006ni}
Pierre Sikivie.
\newblock {Axion cosmology}.
\newblock {\em Lect. Notes Phys.}, 741:19--50, 2008.

\bibitem{strobl:weiler:axion:anharmonic}
Karl Strobl and Thomas~J. Weiler.
\newblock {Anharmonic evolution of the cosmic axion density spectrum}.
\newblock {\em Phys. Rev.}, D50:7690--7702, 1994.

\bibitem{thooft:instanton:fluctuations}
G.~'t~Hooft.
\newblock {Computation of the quantum effects due to a four-dimensional
  pseudoparticle}.
\newblock {\em Phys.Rev. D}, 14:3432--3448, 1976.

\bibitem{turner:axion:cosmology}
M.~Turner.
\newblock {Cosmic and local mass density of invisible axions}.
\newblock {\em Phys. Rev. D}, 33:889, 1986.

\bibitem{turner:axion:report}
M.~Turner.
\newblock {Windows on the Axion}.
\newblock {\em Phys. Rept.}, 197:67--97, 1990.

\bibitem{Vilenkin:1982ks}
A.~Vilenkin and A.~E. Everett.
\newblock {Cosmic Strings and Domain Walls in Models with Goldstone and
  PseudoGoldstone Bosons}.
\newblock {\em Phys. Rev. Lett.}, 48:1867--1870, 1982.

\bibitem{vilenkin:shellard:defects}
A.~Vilenkin and E.P.S. Shellard.
\newblock {\em {Cosmic Strings and Other Topological Defects}}.
\newblock Cambridge {U}niversity {P}ress, 1994.

\bibitem{visinelli:gondolo:axion}
L.~Visinelli and P.~Gondolo.
\newblock {Dark Matter Axions Revisited}.
\newblock {\em Phys. Rev.}, D80:035024, 2009.

\bibitem{wantz:iilm:1}
O.~Wantz.
\newblock {The topological susceptibility from grand canonical simulations in
  the interacting instanton liquid model: zero temperature calibrations and
  numerical framework}.
\newblock {\em Nucl. Phys. B}, 829:48--90, 2010.

\bibitem{wantz:iilm:3}
O.~Wantz and E.P.S. Shellard.
\newblock {The topological susceptibility from grand canonical simulations in
  the interacting instanton liquid model: chiral phase transition and axion
  mass}.
\newblock {\em Nucl. Phys. B}, 829:110--160, 2010.

\bibitem{weinberg:axion}
S.~Weinberg.
\newblock {A new light boson?}
\newblock {\em Phys. Rev. Lett.}, 40:223, 1978.

\bibitem{weinberg:qft}
S.~Weinberg.
\newblock {\em {The quantum theory of fields (Vol 2)}}.
\newblock {Cambridge University Press}, 1996.

\bibitem{wilczek:axion}
F.~Wilczek.
\newblock {Problem of strong $P$ and $T$ invariance in the presence of
  Instantons}.
\newblock {\em Phys. Rev. Lett.}, 40:279, 1978.

\bibitem{yamaguchi:yokoyama:global:strings:lagrangian}
M.~Yamaguchi and J.~Yokoyama.
\newblock {Quantitative evolution of global strings from the Lagrangian view
  point}.
\newblock {\em Phys. Rev. D}, 67:103514, 2003.

\bibitem{zhitnitsky:axion}
A.~Zhitnitsky.
\newblock {On Possible Suppression of the Axion Hadron Interactions. (In
  Russian)}.
\newblock {\em Sov. J. Nucl. Phys.}, 31:260, 1980.

\end{thebibliography}

\end{document}